\documentclass[11pt, a4paper, twoside]{article}		
\usepackage[english]{babel}
\usepackage[T1]{fontenc} 
\usepackage[utf8]{inputenc}
\usepackage{amsmath}
\usepackage{amsfonts}
\usepackage{amssymb} 
\usepackage{lmodern}
\usepackage{jheppub}
\usepackage{amsmath}				
\usepackage{amssymb}
\usepackage{amsthm}				
\theoremstyle{plain}

\usepackage{mathtools}				
\usepackage[]{graphicx}
\usepackage{verbatim}
\usepackage{pgfplots}
\usepackage{physics}
\usepackage{booktabs}
 \usepackage{url}
\usepackage{dsfont}
\usepackage{tikz}
\usepackage{bm}
\usepackage{scalerel}
\usetikzlibrary{arrows,shapes,snakes,automata,backgrounds,petri}
\usetikzlibrary{positioning}

\newcommand{\Li}{\text{Li}}
\usepackage{mathtools}
\tikzstyle{place}=[circle,draw=blue!50,fill=blue!20,thick, minimum width=1.6cm )]
\tikzstyle{transition}=[rectangle,draw=black!50,fill=black!20,thick, minimum width=1.2cm,minimum height=1.2cm]
\tikzstyle{dot}=[dot, draw=black, thick]
\title{Topologically twisted index of $T[SU(N)]$ at large $N$}
\author[a,b]{Lorenzo Coccia}
\affiliation[a]{Dipartimento di Fisica, Università di Milano-Bicocca, I-20126 Milano, Italy}
\affiliation[b]{INFN, Sezione di Milano-Bicocca, I-20126 Milano, Italy}
\emailAdd{l.coccia@campus.unimib.it}

\abstract{We compute, in the large $N$ limit, the topologically twisted index of the 3d $T[SU(N)]$ theory, namely the partition function on $\Sigma_{\mathfrak{g}} \times S^1$, with a topological twist on the Riemann surface $\Sigma_{\mathfrak{g}}$. To provide an expression for this quantity, we take advantage of some recent results obtained for five dimensional quiver gauge theories. In case of a universal twist, we correctly reproduce the entropy of the universal black hole that can be embedded in the holographically dual solution.}

\begin{document}
\maketitle
\newpage
\section{Introduction}
The topologically twisted index  $Z_{\Sigma_\mathfrak{g}\times S^1}$ for three dimensional $\mathcal{N} \ge 2$ gauge theories  \cite{Benini:2015noa,Benini:2016hjo} is the partition function on $\Sigma_{\mathfrak{g}} \times S^1$ with a topological twist on the Riemann surface $\Sigma_{\mathfrak{g}}$. 
Thanks to the insight of \cite{Benini:2015eyy}, the index for the ABJM theory in the large $N$ limit has been used to provide the first holographic microscopic counting of the entropy of an asymptotically AdS black hole. Afterwards, this result has been extended and the large $N$ limit of the index has been studied for other quiver gauge theories with an AdS$_4$ dual \cite{Hosseini:2016tor, Hosseini:2016ume, Jain:2019lqb, Jain:2019euv}. See \cite{Zaffaroni:2019dhb} for a review.

To compute the topologically twisted index, one can resort to the localization results in \cite{Benini:2015noa,Benini:2016hjo} which allow to write the index as a contour integral of a meromorphic form, summed over a lattice of magnetic fluxes. It is convenient, as in \cite{Benini:2015eyy}, to first perform the sum and to apply the residue theorem to evaluate the expression left. In this procedure, one can introduce an auxiliary quantity, the \emph{twisted superpotential}\footnote{This quantity is referred to as \emph{Bethe potential} in \cite{Benini:2015eyy} and part of the literature.}, whose critical points give the position of the poles of the integrand obtained after performing the sum.

Remarkably, in the large $N$ limit of all the studied cases, the twisted superpotential turns out to be related to the free energy on the three-sphere  by a simple relation, pointed out in \cite{Hosseini:2016tor} (see also \cite{Jain:2019lqb}). 
Moreover, in the same limit and for a particular choice of topological twist, called \emph{universal topological twist} in \citep{Azzurli:2017kxo}, the free energy $\mathcal{F}_{S^3}$ is also related to the twisted index of the same theory by
\begin{equation}
\label{verificanda}
\log Z_{\Sigma_\mathfrak{g}\times S^1}=(\mathfrak{g}-1)\mathcal{F}_{S^3} \ . 
\end{equation}
This simple relation has an equally simple explanation, in the holographic perspective, in terms of an \emph{universal black hole} \cite{Azzurli:2017kxo}. This black hole is a solution in minimal four dimensional gauged supergravity  and in the large $N$ limit the topologically twisted index reproduces its entropy. The fact that the black hole solution can be embedded in infinitely many ways in eleven-dimensional and massive type IIA supergravity justifies the fact that \eqref{verificanda} holds for a large class of theories \cite{Azzurli:2017kxo}.  

However, despite all these significant progresses, the index has been computed only for a few theories in the large $N$ limit. First of all, the theories considered so far are non chiral, namely for each bi-fundamental connecting two nodes there is another bi-fundamental in the opposite direction and the number of fundamental and (anti-)fundamental fields is equal. It is not yet clear how to compute the large $N$ limit of the twisted index for chiral quivers.\footnote{This situation is similar to the one found in \cite{Jafferis:2011zi} for the large $N$ limit of the free energy  on $S^3$.}  Moreover, even among non chiral theories, many cases are not covered in the literature. An example is the $T[SU(N)]$ theory.\footnote{We note, however, that the topologically twisted index for $T[SU(N)]$ has been considered in \cite{Crew:2020jyf} from the point of view of the factorisation into holomorphic blocks.}

$T[SU(N)]$ is a three dimensional $\mathcal{N}=4$ gauge theory, originally introduced in the study of S-duality of boundary conditions in $\mathcal{N}=4$ four dimensional SYM theory \cite{Gaiotto:2008ak}. It can be represented in terms of a linear quiver with $N-1$ gauge nodes, as we will review in the next section.
The partition function on $S^3$ of $T[SU(N)]$ has been computed exactly for each $N$ \cite{Benvenuti:2011ga,Nishioka:2011dq} and the leading behaviour of the related free energy $\mathcal{F}_{S^3}$ at large $N$ is
\begin{equation}
\mathcal{F}_{S^3} = \frac{1}{2}N^2 \log N+O(N^2) \ .
\end{equation}
This result has also been reproduced considering the gravity dual solution  of $T[SU(N)]$ \cite{Assel:2011xz,Assel:2012cp}. For our discussion, the important point is that, in line with the general conjecture of  \cite{Gauntlett:2007ma}, we do expect a consistent truncation of this dual solution to $4d$ gauged supergravity. This means that, if we compute the twisted index for $T[SU(N)]$ in case of the universal twist, we expect it to verify relation \eqref{verificanda}, reproducing the entropy of the universal black hole.  This will be one of the aims of this paper.

In \cite{Uhlemann:2019ypp}, the planar limit of the free energy on a five dimensional sphere has been computed for a class of theories, reducing standard saddle point computations in a form akin to a 2d electrostatic problem. Theories considered in \cite{Uhlemann:2019ypp} are described by linear quivers with a large number of nodes. In three dimensions, this is exactly what happens for $T[SU(N)]$ in the large $N$ limit. Indeed, in this work we will show that the approach of \cite{Uhlemann:2019ypp} can also be applied in three dimensions. First of all, we will be able to reproduce the known results for the free energy on the three-sphere, testing in this way the method. Using the same framework, we will then move to the computation of the twisted superpotential, recovering the known relation with the free energy, previously mentioned.
Finally we will also be able to compute the topologically twisted index for $T[SU(N)]$, verifying relation \eqref{verificanda} for the particular choice of the universal topological twist. 

The plan of the rest of the paper is the following. 
In the next section we will introduce the $T[SU(N)]$ theory, recalling some useful aspects and describing the notation used in the paper. In Section \ref{Section_free}, we will explicitly compute the free energy in the large $N$ limit, following the procedure of \cite{Uhlemann:2019ypp}. In this computation we will turn on an arbitrary R-charge, in prevision of a comparison with the expressions for the twisted superpotential and the index, which are functions of chemical potentials. Section \ref{Section_index} is devoted to the review of the general aspects of the topologically twisted index, together with some known results. In Section \ref{Section_bethe} we will compute the twisted superpotential for $T[SU(N)]$ and finally in Section \ref{Section_index_tsun} we will provide the expression for the index, in particular in the case of the universal topological twist. We will conclude the paper with few comments and three appendices, containing useful formulas and computations. 

\section{General features of $T[SU(N)]$}
\label{Section_tsun}
$T[SU(N)]$ is a three dimensional $\mathcal{N}=4$ gauge theory at the IR superconformal fixed point. It was originally introduced in \cite{Gaiotto:2008ak} and admits a nice description in terms of the linear quiver (in $\mathcal{N}=2$ notation):
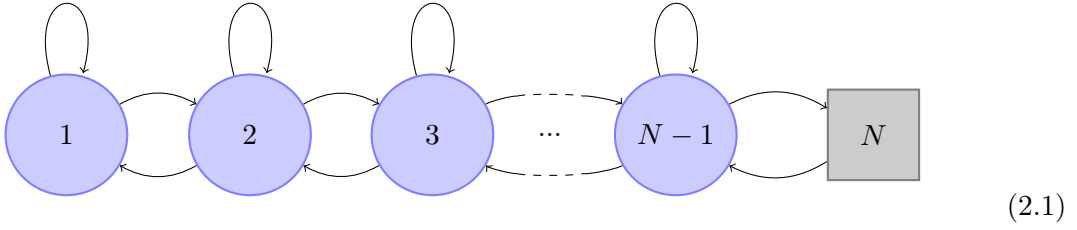
\begin{equation}
\label{quiver_tsun}
\tikzset{
    block/.style={
        rectangle, fill=white!20, 
        minimum width=2.2em, rounded corners, minimum height=4em
        }
    }
\begin{tikzpicture}[inner sep=2mm]
\node[place]   		    (q_0)  						 {$1$};
\node[place]  			(q_1) [right=1.6 cm] 		 {$2$};
\node[place]			(q_2) [right=4.0 cm]  	 	{$3$};
\node[block]			(X) 		[right= 5.93 cm]  	 	{...};
\node[place]			(q_3) [right=7.2 cm]  	 	{$N-1$};
\node[transition]		 (e1) [right=10 cm] {$N$};
\path[->] (q_0) edge [bend left]		node 		{} (q_1);
\path[->] (q_1)	edge [bend left]			node 	    {} (q_2);
\path[<-] (q_0) edge [bend right]		node 		{} (q_1);
\path[<-] (q_1)	edge [bend right]			node 	    {} (q_2)
		  (q_0)	edge [loop above]	node		{} ()
		  (q_1)	edge [loop above]	node		{} ()
		  (q_2)	edge [loop above]	node 		{} ();
\path[<-] (q_3)	edge [bend right]	node 	    {} (e1);
\path[->] (q_3) edge [bend left]	node 		{} (e1)
		  (q_3)	edge [loop above]	node 		{} ();

\draw[<-] (q_2.-30) to[bend right,looseness=.3] (X.230) node[below left=4pt and 5pt] {};
    \draw[dashed] (X.230) to[bend right, looseness=.2] (X.-50); 
    \draw (X.-50) to[bend right, looseness=.3] (q_3.210);

\draw (q_2.30) to[bend left,looseness=.3] (X.130) node[below left=4pt and 5pt] {};
    \draw[dashed] (X.130) to[bend left, looseness=.2] (X.50); 
    \draw[->] (X.50) to[bend left, looseness=.3] (q_3.150);
\end{tikzpicture}
\end{equation}
Each round node in the quiver is labelled by a number $t=1, \dots, N-1$ and denotes a factor $U(t)$ in the gauge group of the theory, with an associated $\mathcal{N}=2$ vector multiplet. On top of each node, an arc indicates the presence of a chiral field $\Phi_t$ in the adjoint representation of the group $U(t)$.
Bi-fundamental fields are identified with lines connecting adjacent nodes: we  denote $Q^{(t)}$ the bi-fundamental which goes from the node $t$ to the node $t+1$ and $\tilde{Q}^{(t)}$ the bi-fundamental in the opposite direction. For the last gauge node, the lines and the related symbols $\tilde{Q}^{(N-1)}, Q^{(N-1)}$ actually denotes (anti-)fundamental fields transforming under the $SU(N)$ global symmetry, represented by the square box at the end of the quiver. With these conventions, the superpotential for $T[SU(N)]$ can be schematically written as
\begin{equation}
\label{superpot}
W=\sum_{t=1}^{N-1} \Tr \left[\tilde{Q}^{(t)}\Phi^{(t)}Q^{(t)}-Q^{(t-1)}\Phi^{(t)}\tilde{Q}^{(t-1)} \right] \ ,
\end{equation}
where $Q^{(0)}=0$ (we refer to \citep{Aprile:2018oau, Hwang:2020wpd} for a more careful notation). 
Possible deformations of the theory are obtained considering mass terms for the $\mathcal{N}=4$ hypermultiplet and Fayet-Iliopoulos (FI) parameters for the $U(1)$ gauge factors. In this work, however, we will set these deformations to zero.

We also recall that $T[SU(N)]$ is invariant under mirror symmetry \cite{Intriligator:1996ex}. This symmetry acts on the vacuum moduli space and exchanges Higgs and Coulomb branches,  mass and FI parameters, and the two factors of the  $SU(2)_R \times SU(2)_L $ R-symmetry of the theory. 

\paragraph{Notation} In the next sections, following the treatment of \cite{Uhlemann:2019ypp}, it will be convenient to rewrite the quiver \eqref{quiver_tsun} in a slightly more general way. We use the symbol $L$ to indicate the length of the quiver and introduce a coordinate $t$ to label the gauge nodes of the quiver, with $t=1, \dots, L$. We also denote the rank of the gauge group in the $t^{\text{th}}$ node with $N_t$ and number of flavours in the last node with $k_L$. The resulting quiver is 
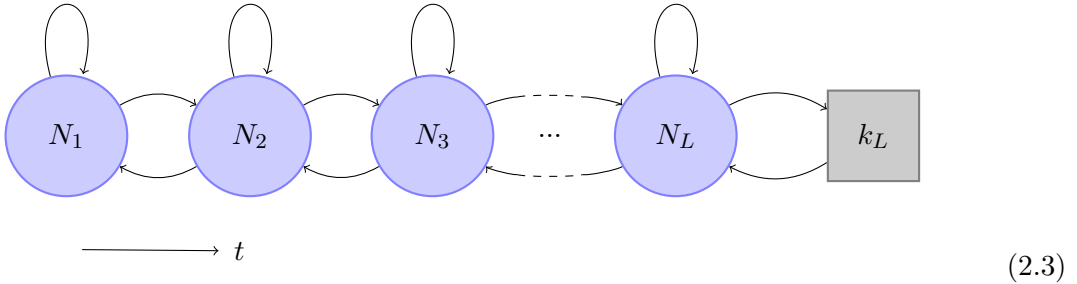
\begin{equation}
\label{quiver_general}
\tikzset{
    block/.style={
        rectangle, fill=white!20, 
        minimum width=2.2em, rounded corners, minimum height=4em
        }
    }
\begin{tikzpicture}[inner sep=2mm]
\node[place]   		    (q_0)  						 {$N_1$};
\node[place]  			(q_1) [right=1.6 cm] 		 {$N_2$};
\node[place]			(q_2) [right=4.0 cm]  	 	{$N_3$};
\node[place]			(q_3) [right=7.2 cm]  	 	{$N_L$};
\node[block]			(X) 		[right= 5.93 cm]  	 	{...};
\node[transition]		 (e1) [right=10 cm] {$k_L$};
\node[]   		    (t_0) [below=1.3 cm]			 {};
\node[]   		    (t_1) [below right=1.2 cm and 2 cm]	 {$t$};
\path[->] (q_0) edge [bend left]		node 		{} (q_1);
\path[->] (q_1)	edge [bend left]			node 	    {} (q_2);
\path[<-] (q_0) edge [bend right]		node 		{} (q_1);
\path[<-] (q_1)	edge [bend right]			node 	    {} (q_2)
		  (q_0)	edge [loop above]	node		{} ()
		  (q_1)	edge [loop above]	node		{} ()
		  (q_2)	edge [loop above]	node 		{} ();
\path[<-] (q_3)	edge [bend right]	node 	    {} (e1);
\path[->] (q_3) edge [bend left]	node 		{} (e1)
		  (q_3)	edge [loop above]	node 		{} ();
\path[->] (t_0) edge []node {} (t_1);
\draw[<-] (q_2.-30) to[bend right,looseness=.3] (X.230) node[below left=4pt and 5pt] {};
    \draw[dashed] (X.230) to[bend right, looseness=.2] (X.-50); 
    \draw (X.-50) to[bend right, looseness=.3] (q_3.210);

\draw (q_2.30) to[bend left,looseness=.3] (X.130) node[below left=4pt and 5pt] {};
    \draw[dashed] (X.130) to[bend left, looseness=.2] (X.50); 
    \draw[->] (X.50) to[bend left, looseness=.3] (q_3.150);

\end{tikzpicture}
\end{equation}
At the end of our computations, we will substitute $N_t=t$, $ L=N-1$ and $k_L=N$, obtaining results for the original quiver \eqref{quiver_tsun}.

\section{Large $N$ free energy on $S^3$}
\label{Section_free}
We start considering the $T[SU(N)]$ partition function on the three dimensional sphere $S^3$. Resorting to the results obtained in \cite{Kapustin:2009kz} using a localization procedure, this quantity has been computed exactly for each $N$ \cite{Benvenuti:2011ga,Nishioka:2011dq}. Moreover, the related free energy 
\begin{equation}
\label{def_free}
\mathcal{F}_{S^3}=-\log \abs{Z_{S^3}} 
\end{equation}
has been studied, in the large $N$ limit, both from the field theory and the gravity dual side\footnote{In \cite{Assel:2012cp,Assel:2011xz} a larger class of $T^{\rho}_\sigma [SU(N)]$ has been considered.} \cite{Assel:2012cp,Assel:2011xz,Assel:2013lpa}. In case of vanishing mass deformations and topological charges, the leading behaviour is
\begin{equation}
\label{asympt_behaviour}
\mathcal{F}_{S^3}=\frac{1}{2}N^2 \log N+O(N^2) \ .
\end{equation}
Our first aim in this work is to take advantage of the recent approach of \cite{Uhlemann:2019ypp} in studying long linear quivers to reproduce the behaviour \eqref{asympt_behaviour}. Indeed, we will see that the techniques applied in \cite{Uhlemann:2019ypp} to compute partition functions on the five-sphere can be re-proposed in three dimensions, providing consistent results. Remarkably, the formalism we will derive in this section for $T[SU(N)]$ can be applied to the free energy of more general three dimensional theories described by long linear quivers \cite{Coccia:2020wtk}. 

\subsection{Large $N$ matrix model}
As shown in \cite{Kapustin:2009kz}, the partition function on the three-sphere of a superconformal theory with $\mathcal{N}=2$ or more supersymmetry localizes to a matrix model. In these cases the R-charges of the theory are fixed to their canonical values. This result has been extended in \citep{Jafferis:2010un, Hama:2010av} where more general $\mathcal{N}=2$ theories with an arbitrary R-charge assignment have been considered. For these theories, the partition function on ${S^3}$ still reduces to a matrix model and it is a function of a set of trial R-charges. Extremizing the partition function returns the exact values for the R-charges \cite{Jafferis:2010un}.

In this section, we write the partition function for $T[SU(N)]$ in language $\mathcal{N}=2$ and turn on an arbitrary R-charge $r$, equal for all the bi-fundamental and (anti-)fundamental fields in the theory. The R-charge $\tilde{r}$ for all the adjoint fields is then fixed by the constraint
\begin{equation}
\label{condition_rcharges}
\tilde{r}=2(1-r) \ ,
\end{equation}
in such a way that the superpotential \eqref{superpot} has R-charge 2. Using the results in \cite{Jafferis:2010un}, we can write the partition function as a matrix model and we will then apply the saddle point approximation to compute the free energy \eqref{def_free} as a function of the charge $r$. At the end, maximizing the expression with respect to $r$, we expect to find \eqref{asympt_behaviour} when $r=1/2$. The choice of turning an R-charge on is motivated by a future comparison of $\mathcal{F}_{S^3}$ with the topological twisted index, where chemical potentials will play the role of R-charges (see Section \ref{sec:rev_results}). Note that, for this reason, we couldn't simply consider the results of \cite{Assel:2012cp,Assel:2011xz,Assel:2013lpa}, where the R-charges have their canonical values.

The partition function we are interested in localizes to a finite matrix integral, which we schematically write \cite{Jafferis:2010un}
\begin{equation}
\label{matrix_mod}
Z_{S^3}=\frac{1}{\abs{\mathfrak{W}}}\int \prod_{\text{Cartan}} d \lambda \ e^{-\mathcal{F}(\lambda)} \ ,
\end{equation} 
where $\abs{\mathfrak{W}}$ is the order of the Weyl group of the gauge group. The exponential in the integrand is the sum of the various  contributions in the theory
\begin{equation}
\label{free_energy_discrete}
\mathcal{F}=F_{\text{vec}}+F_{\text{adj}}+F_{\text{bif}}+F_{\text{(a)f}} \ .
\end{equation} 
More explicitly, and using the notation introduced in the quiver \eqref{quiver_general}, the contribution from all the gauge nodes in the theory is
\begin{equation}
F_{\text{vec}}=  -\sum_{t=1}^L \sum_{i\neq j=1}^{N_t} \frac{1}{2}\log\left(4 \sinh^2(\pi( \lambda_i^{(t)} -\lambda_j^{(t)}))\right)  \ ;
\end{equation}
from the bi-fundamental pairs we have
\begin{equation}
F_{\text{bif}}=-\sum_{t=1}^{L-1} \sum_{i=1}^{N_t}\sum_{j=1}^{N_{t+1}} \left(\ell(1-r+i(\lambda_i^{(t)}-\lambda_j^{(t+1)}))+\ell(1-r-i(\lambda_i^{(t)}-\lambda_j^{(t+1)}))  \right)
\end{equation}
with
\begin{equation}
\ell(z)=-z\log \left(1-e^{2\pi i z}\right)+\frac{i}{2}\left(\pi z^2+\frac{1}{\pi}\text{Li}_2\left(e^{2\pi i z}\right) \right)-\frac{i\pi}{12} 
\end{equation}
and from the (anti-)fundamentals in the last node with $k_L$ flavours
\begin{equation}
F_{\text{(a)f}}= -k_L \sum_{i=1}^{N_L} \left(\ell(1-r+i\lambda_i^{(L)})+\ell(1-r-i\lambda_i^{(L)})  \right) \ . 
\end{equation}
Adjoint fields are identified with a pair of bi-fundamentals connecting the same gauge group, with an overall factor $1/2$, and this gives the expression for $F_{\text{adj}}$.

Our plan is to evaluate \eqref{matrix_mod} in the large $N$ limit, using the saddle point approximation. Hence, the idea is to find the configuration of eigenvalues $\lambda$ extremizing \eqref{free_energy_discrete}. Note that the prefactor $1/\abs{\mathfrak{W}}$ in front of \eqref{matrix_mod} compensates for the fact that there are $\abs{\mathfrak{W}}$ distinct critical points in which the integral takes the same value. Therefore, once we evaluate $F(\lambda)$ in one of such configurations $\lambda_0$, we can approximate the integral \eqref{matrix_mod} as $Z \sim e^{-\mathcal{F}(\lambda_0)}$. It is important to observe that, by this method, we do not expect to be able to reproduce subleading terms of order $N^2$ in the asymptotic expansion. Indeed, the number of integration variables in the matrix model is of order $ N^2$ as well, meaning that all the orders in the expansion of $\mathcal{F}(\lambda)$ around the saddle point could in principle contribute to the order $N^2$ of the free energy.

We now follow the treatment of \cite{Uhlemann:2019ypp}, adapting  it to our case. The extremization problem can be tackled introducing a density for the eigenvalues (i.e.\! the integration variables) of each node
\begin{equation}
\rho_t(\lambda)=\frac{1}{N_t}\sum_{i=1}^{N_t} \delta(\lambda-\lambda_i^{(t)}) \ .
\end{equation}
We assume that, for large $N_t$, this density becomes a continuous function, with the correct normalization
\begin{equation}
\int d\lambda \ \rho_t(\lambda)=1 \ .
\end{equation}
Obviously, a continuous distribution is not a good approximation for the eigenvalues associated with a group of small rank, i.e.\ for the first nodes of $T[SU(N)]$. However, as the quiver becomes longer and longer (i.e.\! for large $N$), we expect the contribution from the large groups to be more and more important and the approximation to be reliable. Hence, in the expression \eqref{free_energy_discrete} we simply substitute
\begin{equation}
\label{discrete_density}
\sum_{i=1}^{N_t} \quad \to \quad N_t \int d\lambda \ \rho_t(\lambda) 
\end{equation}
for each node $t$ obtaining
\begin{equation}
\label{intermediate_free}
\begin{split}
\mathcal{F}= \int d\lambda d\lambda' & \left(\sum_{t=1}^L N_t^2 \rho_{t}(\lambda)\rho_t(\lambda')F_V(\lambda-\lambda')+\sum_{t=1}^{L-1} N_t N_{t+1}\rho_t(\lambda) \rho_{t+1}(\lambda')F_H(\lambda-\lambda') \right)\\
&+k_L\int d\lambda N_L \rho_L (\lambda) F_H(\lambda) \ ,
\end{split}
\end{equation}
where we introduced
\begin{equation}
\label{Fv-Fh}
\begin{split}
F_V(\lambda) &= -\frac{1}{2} \log (4\sinh^2(\pi \lambda))-\frac{1}{2}\left[\ell(1-\tilde{r}+i \lambda)+\ell(1-\tilde{r}-i \lambda)\right] \ , \\
F_H(\lambda) &= -\ell(1-r+i \lambda)- \ell(1-r-i \lambda) \ .
\end{split}
\end{equation}
Here, the first term in $F_V$ is the vector contribution and the second the adjoint one. The remaining bi-fundamental contributions are in $F_H$. After a simple manipulation, \eqref{intermediate_free} becomes 
\begin{equation}
\label{intermediate_free_energy}
\begin{split}
\mathcal{F} & = \int d\lambda d\lambda' \left[\sum_{t=1}^L N_t^2 \rho_t(\lambda) \rho_t(\lambda')F_0(\lambda-\lambda')-\frac{1}{2}\sum_{t=1}^{L-1} \eta_t(\lambda) \eta_{t}(\lambda')F_H(\lambda-\lambda') \right] \\
&-\frac{1}{2}\sum_{t \in \{ 1,L \} } \int d\lambda d\lambda' N_t^2 \rho_t(\lambda) \rho_t(\lambda')F_H(\lambda-\lambda') +k_L\int d\lambda N_L \rho_L (\lambda) F_H(\lambda)
\end{split}
\end{equation}
with $\eta_t (\lambda) \equiv N_{t+1} \rho_{t+1}(\lambda)- N_{t} \rho_{t}(\lambda)$ and 
\begin{equation}
F_0(\lambda)=F_V(\lambda)+F_H(\lambda) \ .
\end{equation}

At this point, the procedure of \cite{Uhlemann:2019ypp} consists in replacing the variable $t$, which labels nodes of the quiver, with the variable $z=t/L$ which can be considered continuous in the large $L$ limit. The boundaries of the quiver are given by $z=0$ and $z=1$. At the same time, we replace the family of densities in  \eqref{discrete_density}, parametrized by $t$, with a single function of two continuous parameters
\begin{equation}
\rho_t(\lambda)=\rho_{zL}(\lambda)\equiv \rho(z,\lambda) \ .
\end{equation}
Promoting the rank of the gauge groups $N_t$ to a continuous function of $z$, we also make the substitution
\begin{equation}
N_{t+1} \rho_{t+1}(\lambda)- N_{t} \rho_{t}(\lambda) \quad \to \quad \frac{1}{L}\partial_z \left(N(z) \rho(z,\lambda) \right) \ .
\end{equation}
We can then rewrite \eqref{intermediate_free_energy} as
\begin{equation}
\label{continuous_free}
\begin{split}
& \mathcal{F}=  L \int_0^1 dz \int d\lambda d\lambda' \Bigg[ N(z)^2\rho(z,\lambda) \rho(z,\lambda')F_0\left(\lambda-\lambda'\right)  \\
& \qquad \qquad \qquad -\frac{1}{2L^{2}}\partial_z \left( N(z) \rho(z,\lambda) \right) \partial_z \left( N(z) \rho(z,\lambda') \right) F_H\left(\lambda-\lambda'\right)\Bigg]  \\
& -\frac{1}{2} \sum_{z \in \{ 0,1 \} } \int d\lambda d\lambda' \ N(z)^2\rho(z,\lambda) \rho(z,\lambda')F_H\left(\lambda-\lambda' \right)+k_L N(1) \int d\lambda  \ \rho (1,\lambda) F_H\left(\lambda\right) \ .
\end{split}
\end{equation}
For $T[SU(N)]$, $N(z)=zL \sim zN$. Expression \eqref{continuous_free} is completely analogous to the expressions found in \cite{Uhlemann:2019ypp} for free energies on the five-sphere.  However, integrands in \eqref{continuous_free} are rather complicated expressions, which need to be simplified. For this purpose, note that the terms inside the square brackets are balanced in $N(z)$ but not in $L$. Therefore, we assume the eigenvalues to scale as
\begin{equation}
\lambda=L^\alpha x 
\end{equation}
with $\alpha >0$. Under this assumption and after some computations described in Appendix \ref{Appenix_free}, we obtain the expressions
\begin{equation}
\label{asymp_f0fh}
\begin{split}
F_0(L^\alpha x)& \sim \frac{2\pi}{L^\alpha}(1-r)r^2 \delta(x)+\dots  \ , \\
F_H(L^\alpha x)& \sim 2\pi  L^\alpha (1-r) \abs{x}+\dots  \ .
\end{split}
\end{equation}
Before proceeding with the computation, let us pause a bit on these functions. We note that when we plug the expression just written for $F_0$ in \eqref{continuous_free}, it produces a local contribution in the eigenvalues. Looking at the computation in Appendix \ref{Appenix_free}, we see that this is true because the leading term in the expansion of $F_0=F_V+F_H$
\begin{equation}
\pi (2-2r-\tilde{r}) N(z)^2\int d\lambda d\lambda' \ \rho(z,\lambda) \rho(z,\lambda') \abs{\lambda-\lambda'}
\end{equation}
vanishes under condition \eqref{condition_rcharges} imposed by the superpotential. This feature is often called \emph{long-range force cancellation}, meaning that the free energy, which is schematically a function of the entire sum $\sum_{i,j}( \lambda_i-\lambda_j)$, only gets contributions from $i \sim j$. However, this is not completely our case, because of the presence of integrals with $F_H$, which is non local.

Let us now go back to the computations. In order to have a non trivial combination between the terms in the square brackets of \eqref{continuous_free}, we require them to have the same scaling with respect to $L$; this leads us to $\alpha=1$. With this choice, the leading order of the free energy \eqref{continuous_free} becomes
\begin{equation}
\label{final_free}
\begin{split}
\mathcal{F}= &  \int_0^1 dz \int dx \Bigg[ \varrho(z,x)^2\tilde{F}_0(r) -\frac{1}{2}\tilde{F}_H (r)\partial_z \varrho(z,x)\int dx' \partial_z \varrho(z,x')\abs{x-x'}\Bigg]  \\
& -\frac{L}{2} \sum_{z \in \{ 0,1 \} } \int dx dx' \varrho(z,x) \varrho(z,x')\tilde{F}_H(r)\abs{x-x'}+L k_L \int dx  \varrho (1,x) \tilde{F}_H(r)\abs{x} \\
&+L\int_0^1 dz \ \mu(z) \left(\int dx  \varrho(z,x)-N(z)\right) 
\end{split}
\end{equation}
where, performing the change of variable $\lambda=L x$, we introduced the rescaled density $dx\varrho(z,x)=d\lambda N(z)\rho(z,\lambda)$ and defined
\begin{equation}
\begin{split}
\tilde{F}_0(r) &= 2 \pi (1-r)r^2 \ , \\
\tilde{F}_H(r) &= 2 \pi (1-r) \ .
\end{split}
\end{equation}
We also inserted a Lagrange multiplier $\mu(z)$ to impose the normalization condition
\begin{equation}
\label{norm_cond}
\int d x \ \varrho(z,x)=N(z) \ .
\end{equation}

\subsection{Saddle point and boundary conditions}
Having the expression \eqref{final_free}, we can proceed with the saddle point approximation and evaluate it around the critical point. Hence, we need to take the variation of \eqref{final_free} w.r.t.\ $\varrho(z,x)$. In this procedure, we get contributions both from the bulk, namely the interior of the interval $z \in [0,1]$, and the boundary\footnote{Recall that, in general, when one has a functional $F[\phi]$ 
\begin{equation*}
F[\phi]=\int_{\mathcal{M}} g(\phi,\nabla \phi) d V+\int_{\partial\mathcal{M}} h(\phi) d \Sigma
\end{equation*}
over a volume $\mathcal{M}$ with boundary $\partial \mathcal{M}$ the variation is given by
\begin{equation*}
\delta F[\phi]= \int_{\mathcal{M}} \left(\frac{\partial g}{\partial \phi}-\nabla \cdot \frac{\partial g}{\partial(\nabla \phi)}\right)\delta\phi d V + \int_{\partial \mathcal{M}} \left(\frac{\partial g}{\partial (\nabla \phi)}\cdot n+\frac{\partial h}{\partial \phi} \right)\delta \phi d \Sigma
\end{equation*}
with $n$ outward-pointing unit vector, normal to the boundary. The two terms have to independently vanish.} made of $z=\{0,1 \}$.
 The functional variation in the interior of $[0,1]$ gives
\begin{equation}
\label{saddle_free_energy}
2\tilde{F}_0(r) \varrho(z,x)+\tilde{F}_H(r)\int dx' \ \partial_z^2 \varrho(z,x')\abs{x-x'}+L\mu(z)=0 \ .
\end{equation}
This equation has to be satisfied for each $x$. In particular, for large $x$, the first term is subleading, since we assume the density to decay at infinity, and \eqref{saddle_free_energy} gives
\begin{equation}
\label{balanced_quiv}
\abs{x}\tilde{F}_H(r)\int dx' \ \partial_z^2 \varrho(z,x')=0 \quad \Rightarrow \quad  \partial^2_z N(z)=0 \ ,
\end{equation}
where we used the normalization condition \eqref{norm_cond}. Eq.\! \eqref{balanced_quiv} is the continuous version of the condition for balanced nodes $2N_t=N_{t-1}+N_{t+1}$, which is certainly satisfied, for each $t$, by $T[SU(N)]$. Following \cite{Uhlemann:2019ypp}, in order to solve \eqref{saddle_free_energy} we consider its second derivative with respect to $x$: 
\begin{equation}
\label{der_saddle_free_energy}
\frac{\tilde{F_0}}{\tilde{F}_H}\partial_x^2 \varrho(z,x)+\partial_z^2 \varrho(z,x)=0
\end{equation}
with
\begin{equation}
\label{ratio_F}
\frac{\tilde{F_0}}{\tilde{F}_H}=r^2 \ .
\end{equation}
Remarkably, if \eqref{der_saddle_free_energy} is solved, then also \eqref{saddle_free_energy} is automatically solved, with vanishing Lagrange multipliers.

\paragraph{Boundary conditions} The boundary contribution in the variation comes from the explicit terms in the second line of \eqref{final_free} and from the derivatives in the first line.  However, due to the normalization condition \eqref{norm_cond}, for $T[SU(N)]$ when $z=0$ we need to have 
\begin{equation}
\label{bound_0}
\varrho(0,x)=0 \ ,
\end{equation}
since $N(z)=zL$. So we only consider the boundary at $z=1$. Assuming vanishing multipliers even on the boundary, the variation gives 
\begin{equation}
-\tilde{F}_H(r)\int dx' \left(\partial_z \varrho(z,x') +L\varrho(z,x')-L k_L \delta(x')\right)\abs{x-x'}\Big\lvert_{z=1}=0 \ .
\end{equation}
For large $L$, the first term in the brackets is subleading and the equation can be satisfied if
\begin{equation}
\label{bound_1}
\varrho(1,x)=k_L \delta(x) \ .
\end{equation}
The saddle point equation \eqref{der_saddle_free_energy}, together with the boundary conditions \eqref{bound_0} and \eqref{bound_1}, defines a two-dimensional "electrostatic" problem, equal to the one found in \cite{Uhlemann:2019ypp} for the 5d $T_N$ theory \cite{Benini:2009gi}. Up to an appropriate rescaling, then, we can directly read the solution obtained in that work\footnote{The useful equations in \cite{Uhlemann:2019ypp} are (3.1), (3.14) and (4.9).} 
\begin{equation}
\label{density}
\varrho_s(z,x)= \frac{k_L \sin(\pi z)}{2r(\cosh(\frac{\pi}{r}x)+\cos(\pi z))}  \ .
\end{equation}
Note that this (rescaled) density is defined on the entire $x$ axis and it is properly normalized. See \cite{Uhlemann:2019ypp} for more details.

\subsection{Evaluation of the free energy}
We can now evaluate $\mathcal{F}$ in the saddle point, i.e.\! we have to substitute in \eqref{final_free} the $\varrho_s(z,x)$ just found. The result gives the required expression for the free energy $\mathcal{F}_{S^3}$. Thanks to the conditions \eqref{bound_0} and \eqref{bound_1}, the explicit boundary terms disappear leaving (after an integration by parts)
\begin{equation}
\begin{split}
\mathcal{F}_{S^3} & = \int_0^1 dz \int d x \ \varrho_s(z,x) \left[\varrho_s(z,x)\tilde{F}_0(r)+\frac{1}{2}\tilde{F}_H(r) \int dx' \ \partial_z^2 \varrho_s(z,x') \abs{x-x'} \right] \\
& - \frac{1}{2}\int_0^1 dz \int dx dx' \tilde{F}_H(r) \partial_z \left[\varrho_s(z,x)\partial_z \varrho_s(z,x') \right]\abs{x-x'}  \ .
\end{split}
\end{equation}
Using the saddle point equation \eqref{saddle_free_energy} with no Lagrange multipliers, the terms inside the square brackets vanish and the expression becomes
\begin{equation}
\label{inte_useful}
\begin{split}
\mathcal{F}_{S^3}  &=-\frac{1}{2}\tilde{F}_H(r) \int_0^1 dz \int dxdx' \partial_z \left[\varrho_s(z,x)\partial_z\varrho_s(z,x') \right]\abs{x-x'}=\\
&=-\frac{1}{2}\tilde{F}_H(r) \int dx dx' \left[\varrho_s(z,x)\partial_z\varrho_s(z,x') \right]^{z=1}_{z=0} \abs{x-x'} \ .
\end{split}
\end{equation}
Finally, using again the boundary conditions \eqref{bound_0} and \eqref{bound_1}, we can reduce this expression to a one dimensional integral
\begin{equation}
\begin{split}
\mathcal{F}_{S^3} = - \frac{k_L}{2} \tilde{F}_H(r)  \int dx  \left[\partial_z \varrho_s(z,x) \right]_{z=1}\abs{x} \ .
\end{split}
\end{equation}
The domain of integration should be the entire real axis but this integral turns out to be divergent in 0. The origin of this divergence is the fact that we have substituted the expressions of $F_0(L x)$ and $F_H(L x)$ with their asymptotic expansions \eqref{asymp_f0fh}, motivated by the fact that $L$ is large.\footnote{Indeed, at least in case of canonical R-charge, it is possible to repeat the whole discussion with the complete expressions of $F_0$ and $F_H$ and check that there are no divergences.} However, this substitution holds until $x$ becomes of the order $\beta/L$ with an arbitrary finite $\beta$. Hence, using the fact that $\varrho(z,x)$ is even in $x$, we introduce a cut-off in the integral  
\begin{equation}
\begin{split}
\mathcal{F}_{S^3} &= \frac{\pi}{2r}\tilde{F}_H(r)k_L^2\int_{\frac{\beta}{L}}^\infty dx \frac{x}{\cosh(\frac{\pi}{r}x)-1}=\frac{\pi}{4r}\tilde{F}_H(r)k_L^2\int_{\frac{\beta}{L}}^\infty dx \frac{x}{\sinh^2(\frac{\pi x}{2r})}=\\
&=\frac{\left(\tilde{F}_0(r)\tilde{F}_H(r)\right)^{1/2}}{\pi}k_L^2\int_{\frac{\beta'}{L}}^\infty dx \frac{x}{\sinh^2(x)}  \ ,
\end{split}
\end{equation}
where we decided, for future convenience, to express everything in terms of $\tilde{F}_0$ and $\tilde{F}_H$, using \eqref{ratio_F}. Solving the last integral at the leading order in $L$ we have
\begin{equation}
\label{F-saddle}
\mathcal{F}_{S^3}= \frac{\left(\tilde{F}_0(r)\tilde{F}_H(r)\right)^{1/2}}{\pi}k_L^2 \log L+ \dots \ .
\end{equation}
Finally, substituting $L=N$ and $k_L=N$ we find the free energy
\begin{equation}
\label{expression_F}
\mathcal{F}_{S^3} = 2r(1-r)N^2 \log N + O(N^2) \ .
\end{equation}
As expected, this quantity has its maximum when $r=1/2$, where its value is exactly \eqref{asympt_behaviour}. Moreover, note that, under mirror symmetry, the R-charge $r$ is sent into $r \to 1-r$ and expression \eqref{expression_F} is invariant under this substitution, consistently with the self-mirror properties of $T[SU(N)]$.
As a final remark, we note that if we consider slightly different quiver theories with $N_t=a t$, $t=1, \dots , L/a$ and $a$ much smaller of the length of the quiver, the previous discussion can be repeated leading to the same result \eqref{expression_F}. This observation is in agreement with \cite{Assel:2012cp}.

\section{Topologically twisted index: general aspects}
\label{Section_index}
The topologically twisted index \cite{Benini:2015noa,Benini:2016hjo} for a $3d$ $\mathcal{N}\ge 2$ theory is defined as the partition function on $\Sigma_{\mathfrak{g}} \times S^1$, with a topological twist on the Riemann surface $\Sigma_{\mathfrak{g}}$ of genus $\mathfrak{g}$. 
The index is expressed in terms of complex fugacities $y$ for the global symmetries and in terms of a set of integer magnetic fluxes $\mathfrak{n}$ on $\Sigma_{\mathfrak{g}}$, parametrizing inequivalent twists. 

Using localizations techniques \cite{Benini:2015noa,Benini:2016hjo,Zaffaroni:2019dhb}, it is possible to reduce the topologically twisted index to a matrix model. Explicitly, for a theory with gauge group $G$, the index is given by 
\begin{equation}
\label{general_Z}
Z_{\Sigma_{\mathfrak{g}} \times S^1}(y,\mathfrak{n)}= \frac{1}{\abs{\mathfrak{W}}}\sum_{\mathfrak{m}\in \Gamma} \oint_{\mathcal{C}} Z_{pert}(\lambda,y,\mathfrak{m},\mathfrak{n}) \left(\det \frac{\partial^2 \log Z_{pert}(\lambda,y,\mathfrak{m},\mathfrak{n})}{\partial iu \partial \mathfrak{m}}\right)^{\mathfrak{g}} \ .
\end{equation}
Here, $\abs{\mathfrak{W}}$ denotes the order of the Weyl group of $G$ and the sum is over magnetic fluxes $\mathfrak{m}$ living in the co-root lattice $\Gamma$ of $G$. The integration is over the zero-mode gauge variable $\lambda^{i(A_t+i \beta \sigma)}$ where $A_t$ is a Wilson line along $S^1$ running over the maximal torus of the gauge group $G$ and $\sigma$ is the real scalar in the vector multiplet running over the corresponding subalgebra. $\beta$ is the radius of $S^1$. In \eqref{general_Z} we also introduced a Cartan-complex valued quantity $u=A_t+i \beta \sigma$, such that $\lambda=e^{iu}$. Lastly, supersymmetry selects a particular contour of integration in \eqref{general_Z}, which can be formulated in terms of the Jeffrey-Kirwan residue. We refer to \citep{Benini:2015noa,Benini:2016hjo} for more details. 

For a theory without Chern-Simons terms and a set of chiral multiplets transforming in representations $\mathfrak{R}_I$ of $G$, the function in the integral \eqref{general_Z} is given by
\begin{equation} \label{eq:int_Z}
Z_{pert} =  \prod_{\alpha \in G}(1-\lambda^{\alpha})^{1-\mathfrak{g}}(i du)^{\text{rank} G} \prod_I\prod_{\rho_I \in \mathfrak{R}_I} \left( \frac{\lambda^{\rho_I/2}y_I^{1/2}}{1-\lambda^{\rho_I}y_I}\right)^{\rho_I(\mathfrak{m})-\mathfrak{n}_I+1-\mathfrak{g}} \ ,
\end{equation}
where $\alpha$ are the roots of $G$, $\rho_I$ are the weights  of the representation $\mathfrak{R}_I$  and we used the notation $\lambda^{\rho}\equiv \lambda^{i\rho(u)}$. We also included the measure of the integrand in this expression. 
Following \cite{Hosseini:2016tor}, we adopted a redundant assignment for fluxes and fugacities: calling $y_f$ the fugacity associated to a flavor symmetry and $\nu_I$ the weight of a chiral field under the same symmetry, we setted $y_I \equiv y_f^{\nu_I}$. Hence, we have the constraint that, for each term $W_a$ in the superpotential of the theory,
\begin{equation}
\label{condition_fug_2}
\prod_{I \in W_a} y_I =1  \ .
\end{equation}
since we require invariance of the superpotential under the global symmetries.
Similarly, in terms of an assignment $\mathfrak{m}^f_a$ for background global symmetries and of the R-charge $r_I$ of a chiral field, we chose
\begin{equation} \label{eq:cond_flux_gen}
	\mathfrak{n}_I \equiv \nu_I(\mathfrak{m}_f)+(1-\mathfrak{g})r_I
\end{equation}
with the constraint 
\begin{equation}
\label{condition_fug}
\sum_{I \in W_a} \mathfrak{n}_I=2(1-\mathfrak{g}) \ , 
\end{equation}
since we want the superpotential to be invariant under the global symmetries and to have charge $2$ under R-symmetry. 
It will also be convenient to introducing chemical potentials $\Delta_I$ such that $y_I=e^{i\Delta_I}$, so that \eqref{condition_fug_2} becomes
\begin{equation}
\label{condition_chem}
\sum_{I \in W_a} \Delta_I \in 2 \pi \mathds{Z} \ .
\end{equation}
In the following, we will take the chemical potentials to be real.

To compute the index, we follow the procedure described in \cite{Benini:2015eyy,Benini:2016hjo}. After interchanging sum and integral in \eqref{general_Z}, we obtain a geometric series. Resumming this series and following the appropriate prescription for the poles, the index can be written as a sum over residues \cite{Benini:2016hjo}
\begin{equation}
\label{splitting_index}
Z_{\Sigma_\mathfrak{g} \times S^1}=\frac{(-1)^{\text{rank}G}}{\abs{\mathfrak{W}}}\sum_{\text{residues}} Z_{\text{pert}} \lvert _{\mathfrak{m}=0} \left(\det \frac{\partial^2 \log Z_{pert}}{ \partial \mathfrak{m}\partial iu}\right)^{\mathfrak{g}-1} 
\end{equation}
where, defining\footnote{As for the free energy on the three-sphere, we introduce two different indices: a superscript running over the different nodes in the quiver theory and a subscript running over the Cartan of the single node.}
\begin{equation}
\label{def_B}
i B^{(a)}_i=\frac{\partial \log Z_{\text{pert}}}{\partial \mathfrak{m}^{(a)}_i} \ ,
\end{equation}
the residues are those satisfying the Bethe ansatz equations (BAEs)\footnote{In fact, we should only keep solutions for which the Vandermonde determinant $\prod_{\alpha \in G} (1-\lambda^\alpha)$ doesn't vanish.}
\begin{equation}
\label{Bethe1}
e^{iB^{(a)}_i}= 1 
\end{equation}
and we will briefly give a more explicit expression for the left hand side of this equation. First, however, note that rewriting \eqref{Bethe1} as $iB^{(a)}_i-2 \pi i n^{(a)}_i =0$ we can conveniently see the solutions of \eqref{Bethe1} as critical points of an appropriate \emph{twisted superpotential}\footnote{This quantity is sometimes referred to as \emph{Bethe potential}.} $\mathcal{W}$. This potential has some ambiguity in its definition and we will stick with the conventions of \cite{Hosseini:2016tor}.

The expression of the topologically twisted index as sum over the critical points of the twisted superpotential has been first derived in the contest of the Bethe/gauge correspondence, see \cite{Okuda:2012nx,Okuda:2013fea,Okuda:2015yea} and the general discussions in \cite{Nekrasov:2009uh,Nekrasov:2014xaa}. From this perspective, $\mathcal{W}$ is interpreted as the twisted superpotential of the two dimensional theory obtained after the compactification of the 3d theory on $S^1$ \cite{Nekrasov:2009uh,Nekrasov:2014xaa}. See also \cite{Zaffaroni:2019dhb,Gukov:2015sna,Closset:2016arn,Closset:2017zgf,Closset:2017bse,Closset:2018ghr}. 

\subsection{Twisted superpotential} As said, the twisted superpotential is such that its critical points satisfy equation \eqref{Bethe1}. The explicit expression for $B_i^{(a)}$ is \cite{Hosseini:2016tor}
\begin{equation}
\label{B-expr}
\begin{split}
e^{iB_i^{(a)}}= \prod_{\substack{\text{bi-fundamentals} \\ (a,b) \ \text{and} \ (b,a)}} \prod_{j=1}^{N_b} \frac{\sqrt{\frac{\lambda_i^{(a)}}{\lambda_j^{(b)}}y_{(a,b)}}}{1-\frac{\lambda_i^{(a)}}{\lambda_j^{(b)}}y_{(a,b)}}\frac{1-\frac{\lambda_j^{(b)}}{\lambda_i^{(a)}}y_{(b,a)}}{\sqrt{\frac{\lambda_j^{(b)}}{\lambda_i^{(a)}}y_{(b,a)}}} \prod_{\substack{\text{fund.} \\ a}}\frac{\sqrt{\lambda_i^{(a)}y_a}}{1-\lambda_i^{(a)}y_a}\prod_{\substack{\text{anti.} \\ a}}\frac{1-\frac{1}{\lambda_i^{(a)}}\hat{y}_a}{\sqrt{\frac{1}{\lambda_i^{(a)}}\hat{y}_a}} \ ,
\end{split}
\end{equation}
where the different terms coming from bi-fundamental and (anti-)fundamental fields can be identified. Adjoints can be thought as bi-fundamentals connecting the same group. 

After few manipulations of \eqref{B-expr}, one can find the different contributions to the twisted superpotential. Explicitly, a pair of bi-fundamentals, one with chemical potential $\Delta_{(a,b)}$ transforming in the $(\mathbf{N}_a,\bar{\mathbf{N}}_b)$ of $U(N_a) \times U(N_b)$  and the other with chemical potential $\Delta_{(b,a)}$ and transforming in the $(\bar{\mathbf{N}}_a,\mathbf{N}_b)$ of the same group, gives a contribution \cite{Hosseini:2016tor,Jain:2019lqb}
\begin{equation}
\label{building_block1}
\begin{split}
\mathcal{W}^{\text{bi-fund}}= &\sum_{\substack{\text{bi-fundamentals} \\ (a,b) \ \text{and} \ (b,a)}}  \sum_{i=1}^{N_a}\sum_{j=1}^{N_b}\left[\text{Li}_2 \left(e^{i(u_j^{(b)}-u_i^{(a)}+\Delta_{(b,a)})} \right)-\text{Li}_2 \left(e^{i(u_j^{(b)}-u_i^{(a)}-\Delta_{(a,b})}\right) \right] \\
&- \sum_{\substack{\text{bi-fundamentals} \\ (a,b) \ \text{and} \ (b,a)}} \sum_{i=1}^{N_a}\sum_{j=1}^{N_b}\left[\frac{(\Delta_{(b,a)}-\pi)+(\Delta_{(a,b)}-\pi)}{2}(u_j^{(b)}-u_i^{(a)}) \right] \ ,
\end{split}
\end{equation}
where we used $\lambda= e^{iu}$. Similarly, the (anti)-fundamentals contribution is 
\begin{equation}
\begin{split}  
\label{building_block2}
\mathcal{W}^{\text{(anti-)fund}}= & \sum_{i=1}^{N_a}\left[\sum_{\text{anti.} \ a} \text{Li}_2 \left(e^{i(-u_i^{(a)}+\hat{\Delta}_{a})} \right)-\sum_{\text{fund.} \ a} \text{Li}_2 \left(e^{i(-u_i^{(a)}-\Delta_{a})}\right) \right] \\
&+\frac{1}{2} \sum_{i=1}^{N_a} \left[ \sum_{\text{anti.} \ a} (\hat{\Delta}_{a}-\pi)u_i^{(a)}+\sum_{\text{fund.} \ a}  (\Delta_{a}-\pi)u_i^{(a)} \right] \\
& -\frac{1}{4} \sum_{i=1}^{N_a}\left[\sum_{\text{anti.} \ a}  \left( u_i^{(a)}\right)^2-\sum_{\text{fund.} \ a}  \left(u_i^{(a)}\right)^2 \right] \ .
\end{split}
\end{equation}
Note that, for non chiral quivers like $T[SU(N)]$, the last line of this expression vanishes.
\subsection{Brief review of known results} \label{sec:rev_results}As mentioned in the Introduction, in the large $N$ limit the twisted index has been computed for many $\mathcal{N}\ge 2$  quiver gauge theories with M-theory or massive type IIA duals \cite{Hosseini:2016tor, Hosseini:2016ume, Jain:2019lqb, Jain:2019euv}. This subsection is then devoted to a review of the main results found for large $N$. In all the studied cases, under the condition
\begin{equation}
\label{rel_delta}
\sum_{I \in W_a} \Delta_I = 2\pi \ ,
\end{equation}
the twisted superpotential and the free energy on $S^3$ of the same theory are related by the relation, pointed out in \cite{Hosseini:2016tor},
\begin{equation}
\label{bethe-free}
-\frac{2i}{\pi}\widetilde{\mathcal{W}}(\Delta_I)=\mathcal{F}_{S^3}\left(\frac{\Delta_I}{\pi}\right) \ ,
\end{equation}
where with $\widetilde{\mathcal{W}}$ we denote the extremal value of twisted superpotential with respect to the eigenvalues $u$.
It is interesting to observe that this equation relates the free energy to an apparently auxiliary quantity. Moreover, the chemical potentials, which are angular variables, play the role of the R-charges. However, recall that the chemical potentials are constrained by the superpotential and, under the condition \eqref{rel_delta}, they can be safely identified with a set of R-charges.

Another remarkable result is the \emph{index theorem} introduced in \cite{Hosseini:2016tor} (see also \cite{Jain:2019lqb}), which relates twisted superpotential and topologically twisted index of the same theory through 
\begin{equation} \
\label{index-theorem}
\log Z_{\Sigma_\mathfrak{g}\times S^1}(\Delta_I,\mathfrak{n}_I)=(1-\mathfrak{g})\left(\frac{2i}{\pi}\widetilde{\mathcal{W}}(\Delta_I)+i\sum_I \left[\left(\frac{\mathfrak{n}_I}{1-\mathfrak{g}}-\frac{\Delta_I}{\pi} \right)\frac{\partial \widetilde{\mathcal{W}}(\Delta_I)}{\partial \Delta_I} \right] \right) \ .
\end{equation}

As said, for a fixed Riemann surface $\Sigma_{\mathfrak{g}}$, different choices of fluxes parametrize different topological twists. If, in particular, one chooses the fluxes to be proportional to the exact R-charges $\bar{\Delta}_I$ of the theory
 \begin{equation}
 \label{universal_twist}
\bar{\mathfrak{n}}_I=\frac{\bar{\Delta}_I}{\pi}(1-\mathfrak{g})
\end{equation}
one obtains
\begin{equation}
\label{univ-index-free}
\log Z_{\Sigma_\mathfrak{g}\times S^1}(\bar{\Delta}_I,\bar{\mathfrak{n}}_I)=(\mathfrak{g}-1)\mathcal{F}_{S^3}\left(\frac{\bar{\Delta}_I}{\pi}\right) \ . 
\end{equation}
The choice \eqref{universal_twist} is referred to as \emph{universal twist} in \cite{Azzurli:2017kxo} where the authors provide a nice holographic interpretation of \eqref{univ-index-free} in terms of the magnetically charged black hole of \cite{Romans:1991nq,Caldarelli:1998hg}. Being a solution of minimal gauged supergravity, this black hole can be embedded  in eleven dimension and massive type IIA supergravity in infinitely many ways, providing a simple and unique explanation of \eqref{univ-index-free}.
  
We can now go back to the evaluation of the topological twisted index for $T[SU(N)]$. Since, as raised in the Introduction, we expect that the universal black hole can be embedded in the holographically dual solution of $T[SU(N)]$, we also expect the relation \eqref{univ-index-free} to hold even in the $T[SU(N)]$ case. We will verify that this is actually true. Moreover, as an intermediate step in the computation, we will obtain an explicit expression for the twisted superpotential, also verifying \eqref{bethe-free} and \eqref{index-theorem}.

\section{$T[SU(N)]$ twisted superpotential}
\label{Section_bethe}
The discussion of the previous section has been done in the $\mathcal{N}=2$ formalism. Studying an $\mathcal{N}=4$ theory, like $T[SU(N)]$, some observations are in order \cite{Closset:2016arn}. Indeed, starting from the $\mathcal{N}=4$ $R$-symmetry $SU(2)_H \times SU(2)_C$, we can consider different choices of the $\mathcal{N}=2$  R-symmetry  $U(1)$ and perform different topological twists. Once the twist is performed, we are left with an additional global $U(1)$ symmetry with an associated fugacity and a flux (see \cite{Closset:2016arn} for a discussion). Using the conventions introduced in section \ref{Section_index}, we associate the same flux $\mathfrak{n}_h$ to every (anti-)fundamental and bi-fundamental field, and a flux $\mathfrak{n}_v$ to each adjoint field, with the constraint
\begin{equation}
\label{cond_fluxes}
2\mathfrak{n}_h+\mathfrak{n}_v=2(1-\mathfrak{g}) \ .
\end{equation}
Note that we correctly end up with only one free parameter. Moreover, since our fluxes are generic, we are not choosing a particular R-charge in \eqref{eq:cond_flux_gen} and so we do not restrict to a particular twist.
We also assume that every (anti-)fundamental and bi-fundamental field has chemical potential $\Delta$ and every adjoint field a chemical potential $\tilde{\Delta}$. We fix the angular ambiguity requiring that $0 < \Delta,\tilde{\Delta} < 2\pi$ and we choose
\begin{equation}
\label{rel_delta2}
2\Delta+\tilde{\Delta}=2\pi 
\end{equation}
to satisfy the constraint \eqref{condition_chem}. In principle, we could also associate other fugacities and fluxes to the flavor group in  the last node of the $T[SU(N)]$ quiver. However, those contributions would be subleading, as we discuss below equations \eqref{contr_antif_2} and \eqref{index_fund}.

With these conventions, we can apply the rules shown in the previous section to the $T[SU(N)]$ case. Using again the notation of the quiver \eqref{quiver_general}, we write the twisted superpotential as
\begin{equation}
\label{discrete_bethe_0}
\mathcal{W}=  \sum_{t=1}^L \sum_{i, j=1}^{N_t} V_A(u_j^{(t)} -u_i^{(t)})+\sum_{t=1}^{L-1} \sum_{j=1}^{N_{t+1}}\sum_{i=1}^{N_{t}} V_H(u_j^{(t+1)}-u_i^{(t)})+ k_L \sum_{i=1}^{N_L} V_H (-u_i^{(L)}) \ ,
\end{equation}
where 
\begin{equation}
\label{VH}
\begin{split}
V_H\left(x\right)= &\text{Li}_2 \left(e^{i(x+\Delta)} \right)-\text{Li}_2 \left(e^{i(x-\Delta)}\right)-(\Delta-\pi)x \ , \\
V_A\left(x\right)= & \frac{1}{2}\Bigg[ \text{Li}_2 \left(e^{i(x+\tilde{\Delta})} \right)-\text{Li}_2 \left(e^{i(x-\tilde{\Delta})}\right)-(\tilde{\Delta}-\pi)x \Bigg] \ .
\end{split}
\end{equation}
The first term in \eqref{discrete_bethe_0} represents the contribution from the adjoints, the second from the bi-fundamentals and the third from the (anti-)fundamentals in the last node. Note that there is no contribution to the twisted superpotential from the $\mathcal{N}=2$ vector multiplet.

\paragraph{Large $N$ limit} Our first task is to manipulate the  expression \eqref{discrete_bethe_0} and put it, after the long quiver limit (namely the large $N$ limit), in a form analogous to \eqref{final_free}. First of all, we rewrite \eqref{discrete_bethe_0} as
\begin{equation}
\label{discrete_bethe}
\begin{split}
&\mathcal{W} = \sum_{t=1}^L \sum_{i,j=1}^{N_t} \left(V_A(u_j^{(t)} -u_i^{(t)})+V_H(u_j^{(t)} - u_i^{(t)})\right) \\
& -\frac{1}{2}\sum_{t=1}^{L-1}\left[\sum_{i,j=1}^{N_{t+1}} V_H(u_j^{(t+1)}-u_i^{(t+1)})-2\sum_{j=1}^{N_{t+1}}\sum_{i=1}^{N_{t}}V_H(u_j^{(t+1)}-u_i^{(t)})+\sum_{i,j=1}^{N_{t}}V_H(u_j^{(t)}-u_i^{(t)}) \right] \\
&-\frac{1}{2}\sum_{t \in \{ 1, L \}} \sum_{i,j=1}^{N_{t}}V_H(u_j^{(t)}-u_i^{(t)}) + k_L \sum_{i=1}^{N_L} V_H (-u_i^{(L)}) \ .
\end{split}
\end{equation}
We suppose the eigenvalues $u$ to be pure imaginary and to scale with the length $L$ of the quiver according 
\begin{equation}
u_j^{(t)}=i L^\alpha x_j^{(t)}
\end{equation}
 with $x_j^{(t)}$ real. We will fix $\alpha$ later. As for the free energy, we introduce a density for each node $t$ 
\begin{equation}
\sum_{i=1}^{N_t}f(x_i^{(t)}) \ \rightarrow \ \int d x \  \varrho_t(x) f(x) \ , \qquad \qquad \int d x \ \varrho_t(x)=N_t 
\end{equation}
and, in the limit of large $L$, we can simplify the various contributions in \eqref{discrete_bethe}. Consider for example
\begin{equation}
\label{contr_bif}
\sum_{i=1}^{N_q}\sum_{j=1}^{N_p}V_H(u_j^{(p)} - u_i^{(q)}) \ .
\end{equation}
This term is the contribution to the twisted superpotential from a pair of bi-fundamental fields. In case of variables scaling with a large parameter, in our case $L^\alpha$, its expression can be obtained from the results in \cite{Hosseini:2016tor,Jain:2019lqb}. Following the conventions of \cite{Hosseini:2016tor}, we have a local term\footnote{The equations of \cite{Hosseini:2016tor} one has to look at are (A.24) and (A.28), with $\delta v(t)=0$ and with a straightforward generalization to the case of gauge groups with different densities. In \cite{Hosseini:2016tor} the eigenvalues are supposed to scale with $N^{\alpha}$ and the local bi-fundamental contribution scales as $N^{2-\alpha}$; in our \eqref{bif_bethe} a factor $N(z)^2$ is hidden in the densities. See also \cite{Jain:2019lqb} where gauge groups with different ranks are considered.}
\begin{equation}
\label{bif_bethe}
2iL^{-\alpha} \int dx \varrho_p(x)\varrho_q(x) \ g_+(\Delta) 
\end{equation}
plus a non local term 
\begin{equation}
\label{bif_bethe_nonloc}
(\Delta-\pi)\sum_{i \neq j}(u_j^{(p)}-u_i^{(q)})\text{sign}(i-j)  \quad \to \quad  -iL^{\alpha}(\Delta-\pi)\int dx dx' \varrho_p(x)\varrho_q(x') \abs{x'-x} \ .
\end{equation}
In \eqref{bif_bethe} we introduced the function
\begin{equation}
\label{def_g}
g_{+}(u)=\frac{u^3}{6}- \frac{\pi}{2}u^2+\frac{\pi^2}{3}u \ .
\end{equation}
The contribution from adjoint fields, which in the expression \eqref{discrete_bethe} corresponds to the terms with $V_A$, is simply obtained considering $p=q$, with an overall factor $1/2$. For the (anti-)fundamentals, instead,
\begin{equation}
\label{contr_antif}
\sum_{j=1}^{N_p}V_H(- u_i^{(p)})
\end{equation}
the leading order is the contribution
\begin{equation}
\label{contr_antif_2}
-iL^{\alpha}(\Delta-\pi) \int dx \varrho_p(x) \abs{x} \ .
\end{equation}
 For more details on similar computations we refer to \cite{Hosseini:2016tor, Jain:2019lqb}. Note that, if we associate other fugacities to the fundamentals and anti-fundamentals in the last node, the corresponding chemical potentials, $\Delta_f$ and $\Delta'_f$, must satisfy the constraint $\Delta_f =-\Delta'_f$. Hence, their contribution vanishes in equation \eqref{contr_antif_2}. 

\paragraph{Manipulations} We can now use the results reviewed in the previous paragraph to write \eqref{discrete_bethe} in a more convenient form. Consider the first line for fixed $t$, namely
\begin{equation}
\sum_{i,j=1}^{N_t} \left(V_A(u_j^{(t)} -u_i^{(t)})+V_H(u_j^{(t)} - u_i^{(t)})\right) \ .
\end{equation}
This term is the sum of the contribution of an adjoint and a pair of bi-fundamentals connecting the same node. Looking at \eqref{bif_bethe}, then, we see that it gives
\begin{equation}
\begin{split}
\label{first_bethe}
2iL^{-\alpha} \int dx  \varrho_t(x)^2 g_+& \left(\Delta \right)+iL^{-\alpha} g_+(\tilde{\Delta})\int dx \varrho_t(x)^2=iL^{-\alpha} \Delta^2(\pi-\Delta)\int dx \ \varrho_t (x)^2\ ,
\end{split}
\end{equation}
where we used the definition \eqref{def_g} of $g_+(u)$ and \eqref{rel_delta2}. Together with this term, local in the density, the first line also produces a long-range contribution (see eq.\ \eqref{bif_bethe_nonloc}) 
\begin{equation}
\label{non_loc}
\frac{(2\Delta+\tilde{\Delta}-3\pi)}{2}\sum_{i \neq j} \left(u_j^{(t)}-u_i^{(t)}\right)\text{sign}(i-j)\ .
\end{equation}
However, in the definition of the twisted superpotential there is an angular ambiguity and, for each node, we can add a term
\begin{equation}
\label{ambiguity}
-2 \pi \sum_{i=1}^{N_t} n_i^{(t)}u_i^{(t)} 
\end{equation}
with $n_i^{(t)}$ integer. Following \cite{Hosseini:2016tor}, we can use this term to absorb\footnote{In fact, this is actually true only when the dimension of the rank of the gauge group is odd. When we are considering a node with even dimension, instead, we have to include an extra $(-1)^{\mathfrak{m}}$ in the twisted partition function which can be reabsorbed in the definition of the topological fugacity. See \cite{Hosseini:2016tor,Jain:2019lqb} and especially \cite{Closset:2017zgf} for a discussion on sign ambiguities.} the non local contribution \eqref{non_loc}, once one imposes the superpotential constraint $2\Delta+\tilde{\Delta}=2\pi \mathds{Z}$. Note that, in the free energy computation, something similar happens for the term $F_0=F_V+F_H$. There, the long-range forces from vector, adjoint and bi-fundamental fields are perfectly balanced when the R-charges satisfy $2r+\tilde{r}=2$. For the twisted superpotential case, the vector contribution is absent but we can use the ambiguity \eqref{ambiguity}.  To summarize, from the first line of \eqref{discrete_bethe} we only have the contribution given by \eqref{first_bethe}, summed over the nodes.

Consider now the second line
 \begin{equation*}
 -\frac{1}{2}\sum_{t=1}^{L-1}\left[\sum_{j=1}^{N_{t+1}}\sum_{i=1}^{N_{t+1}}V_H(u_j^{(t+1)}-u_i^{(t+1)})-2\sum_{j=1}^{N_{t+1}}\sum_{i=1}^{N_{t}}V_H(u_j^{(t+1)}-u_i^{(t)})+\sum_{j=1}^{N_{t}}\sum_{i=1}^{N_{t}}V_H(u_j^{(t)}-u_i^{(t)}) \right] \ .
 \end{equation*}
In the planar limit, the leading order is given by the term
\begin{equation}
\label{second_bethe}
\begin{split}
\frac{iL^{\alpha}\left(\Delta-\pi \right)}{2}\sum_{t=1}^{L-1}\int dx dx' \abs{x-x'} \left(\varrho_{t+1}(x)-\varrho_{t}(x)\right)\left(\varrho_{t+1}(x')-\varrho_{t}(x')  \right) \ .
\end{split}
\end{equation}
Note that in this case we remain with a long force contribution, as for the computation for the free energy on the three sphere. 

In the end, in the third line of \eqref{discrete_bethe}, we have the boundary terms
\begin{equation}
\frac{i}{2}L^{\alpha}\left(\Delta-\pi \right) \sum_{t \in \{ 1, L \}} \int dx dx' \abs{x-x'}\varrho_t(x)\varrho_t(x') -iL^\alpha k_L \int dx \varrho_L(x) \abs{x}(\Delta-\pi) \ .
\end{equation}

\paragraph{Evaluation of $\widetilde{\mathcal{W}}$} All in all, the twisted superpotential in \eqref{discrete_bethe},  at the leading order in $L$, becomes
\begin{equation}
\label{ev_W}
\begin{split}
\frac{\mathcal{W}}{i}=& L^{-\alpha} \tilde{V}_0(\Delta)\sum_{t=1}^{L}  \int dx \varrho_t (x)^2-\frac{L^{\alpha}}{2} \tilde{V}_H(\Delta)\sum_{t=1}^{L-1}\int dx dx' \abs{x-x'}\eta_{t}(x)\eta_{t+1}(x') \\
&-\frac{L^{\alpha}}{2}\sum_{t \in \{ 1, L \}}\tilde{V}_H(\Delta)\int dx dx' \abs{x-x'}\varrho_t(x)\varrho_t(x')+L^\alpha k_L \tilde{V}_H(\Delta)\int dx \varrho_L(x) \abs{x}
\end{split}
\end{equation}
with $\eta_t(x)=\varrho_{t+1}(x)-\varrho_t(x)$ and
\begin{equation}
\begin{split}
\tilde{V}_0(\Delta) &=\Delta^2(\pi-\Delta) \ , \\
\tilde{V}_H(\Delta) &=\pi-\Delta \ .
\end{split}
\end{equation}
In order to balance the first two terms in \eqref{ev_W} and find the value of $\alpha$, we consider the continuum limit for the variable $t$, as we did for the free energy
\begin{equation}
z=\frac{t}{L} \ , \qquad \varrho(z,\lambda)\equiv \varrho_{zL}(\lambda) \ .
\end{equation}
We obtain
\begin{equation}
\begin{split}
\label{B1}
\frac{\mathcal{W}}{i}=& \int_0^1 dz\int dx \Bigg[L^{1-\alpha} \varrho(z,x)^2 \tilde{V}_0(\Delta)-\frac{L^{\alpha-1}}{2}\tilde{V}_H(\Delta)\int dx' \partial_z\varrho(z,x) \partial_z \varrho(z,x')  \abs{x-x'} \Bigg] \\
&-\frac{L^\alpha}{2}\sum_{z \in \{ 0, 1 \} } \tilde V_H(\Delta) \int dx dx' \varrho(z,x) \varrho(z,x')\abs{x-x'} +L^{\alpha} k_L\tilde{V}_H(\Delta) \int dx \ \varrho(1,x) \abs{x} \ .
\end{split}
\end{equation}
We want to balance the terms in the square brackets and this leads to $\alpha=1$. 
The expression we find with this choice is equal to the expression \eqref{final_free} found for the free energy, up to the substitutions
\begin{equation}
\label{substitution}
\begin{split}
 \tilde{V}_0(\Delta) \ &\to \ \tilde{F}_0(r)  \ ,\\
\tilde{V}_H(\Delta) \ &\to \ \tilde{F}_H(r)  
\end{split}
\end{equation}
and the adding of Lagrange multipliers. Therefore, we can avoid to repeat the whole discussion of Section \ref{Section_free} and immediately write the results. In particular, the saddle point equation for the twisted superpotential is
\begin{equation}
2\tilde{V}_0(\Delta) \varrho(z,x)+\tilde{V}_H(\Delta)\int \ dx' \partial_z^2 \varrho(z,x')\abs{x-x'}+L\mu(z)=0 
\end{equation}
with boundary conditions
\begin{equation}
\label{bound_W}
\varrho(0,x)=0 \ , \qquad \varrho(1,x)=k_L \delta(x) \ ,
\end{equation}
satisfied, for vanishing Lagrange multipliers, by the (rescaled) density
\begin{equation}
\label{density2}
\varrho_s(z,x)= \frac{k_L \sin(\pi z)}{2\Delta(\cosh(\frac{\pi}{\Delta}x)+\cos(\pi z))}  \ .
\end{equation}
Moreover, starting from \eqref{F-saddle} and using the substitution \eqref{substitution}, we can immediately write down the expression for the twisted superpotential in the saddle point
\begin{equation}
\label{value_bethe}
\widetilde{\mathcal{W}}(\Delta)=i \frac{\Delta(\pi-\Delta)}{\pi}N^2 \log N + O(N^2) \ .
\end{equation}
Finally, comparing this expression with \eqref{expression_F} we find the relation
\begin{equation}
-\frac{2i}{\pi}\widetilde{\mathcal{W}}(\Delta)=\mathcal{F}_{S^3}\left(\frac{\Delta}{\pi}\right) \ ,
\end{equation}
as expected.

\section{Topologically twisted index of $T[SU(N)]$}
\label{Section_index_tsun}
We now go back to the evaluation of the topologically twisted index. In Section \ref{Section_index}, we noted that the index can be written as
\begin{equation}
\label{zz}
Z_{\Sigma_\mathfrak{g} \times S^1}=\frac{(-1)^{\text{rank} G}}{\abs{\mathfrak{W}}}\sum_{\text{BEAs}} Z_{\text{pert}} \lvert _{\mathfrak{m}=0} \left(\det \frac{\partial^2 \log Z_{pert}}{\partial iu \partial \mathfrak{m}}\right)^{\mathfrak{g}-1} 
\end{equation}
and we are now going to consider the logarithm of this expression. Note that the physically unambiguous quantity in the planar limit is the real part of the logarithm of the index; so we ignore the overall phase in  \eqref{zz}. 

First of all, let us evaluate the determinant. Recalling the definition \eqref{def_B} and how we introduced the twisted superpotential, we can also write
\begin{equation}
\det \left(\frac{\partial^2 \log Z_{pert}}{\partial iu \partial \mathfrak{m}}\right)=\det \left(\frac{\partial^2 \mathcal{W}}{\partial u \partial u'}\right)\equiv \det \mathds{B} \ .
\end{equation}
Suppose all the entries of $\mathds{B}$ to be bounded\footnote{This assumption can be explicitly verified taking the derivatives of the building blocks \eqref{building_block1} and \eqref{building_block2}: the outcome is that divergences can only occur in regions where $u$ is equal or opposite to the chemical potential $\Delta$. This however is not possible since, with our assumptions, $u$ is pure imaginary and the chemical potential real and non vanishing. A similar discussion on the determinant has been previously proposed in \cite{Benini:2015eyy}. In that work $u$ also has an imaginary part and it is necessary to consider some "tail contributions" related to divergent entries in the matrix.} by some constant $c$. In general, we have that the $t^{\text{th}}$ node's contribution to the matrix is made of $N_t$ lines with $N_{t-1}+N_{t+1}+N_{t}$ non-vanishing entries, coming from bi-fundamentals and adjoint terms. So
\begin{equation}
\label{det_contr}
\log \det \left(\frac{\partial^2 \mathcal{W}}{\partial u \partial u'}\right) \le \log \left( \prod_{t=1}^L c^{N_t}  N_t (N_{t-1}+N_t+N_{t+1} )\right) \ .
\end{equation} 
Since for $T[SU(N)]$ we have $N_t=t$ and $L= N-1$, we can also write
\begin{equation}
\label{det_contr}
\log \det \left(\frac{\partial^2 \mathcal{W}}{\partial u \partial u'}\right) \le \log \left(c^{N^2} \prod_{t=1}^N 3N^2 \right) \sim N^2 .
\end{equation} 
Hence, the contribution from the determinant in \eqref{zz} is at most of order $N^2$ and, we shall see, is subleading in the computation of the index. 

Consider now the remaining part of the index, namely $\log Z_{\text{pert}}$; to simplify the notation, we will omit the subscript. After some manipulations shown in Appendix \ref{Appendix_index}, we can write
\begin{equation}
\begin{split}
\log Z=&\sum_{t=1}^L \sum_{i\neq j=1}^{N_t} (\mathfrak{g}-1)\text{Li}_1 \! \left(e^{i(u_j^{(t)}-u_i^{(t)})} \right)+\sum_{t=1}^L \sum_{i,j=1}^{N_t} Z_A \left(u_j^{(t)}-u_i^{(t)} \right)  \\
& +\sum_{t=1}^L \sum_{i=1}^{N_t}\sum_{j=1}^{N_{t+1}}Z_H\left(u_j^{(t+1)}-u_i^{(t)}\right)+  k_L \sum_{i=1}^{N_L} Z_H \left( u_i^{(L)} \right)
\end{split}
\end{equation}
where the first term is the vector contribution, and the adjoint and bi-fundamentals contributions are respectively expressed in terms of the functions
\begin{equation}
\begin{split}
Z_A(x)& =\frac{1}{2}(-\mathfrak{n}_v+1-\mathfrak{g})\left[\text{Li}_1 \left(e^{i(x+\tilde{\Delta})}\right)+\text{Li}_1 \left(e^{i(x-\tilde{\Delta})}\right)+i(x-\pi) \right] \ , \\
Z_H(x) &= (-\mathfrak{n}_h+1-\mathfrak{g}) \left[\text{Li}_1\left(e^{i\left(x-\Delta \right)}\right)+\text{Li}_1 \left( e^{i\left(x+\Delta \right)} \right)+i(x-\pi) \right] \ . 
\end{split}
\end{equation}
By analogy with the computation of the twisted superpotential, we write
\begin{equation}
\label{trick_for_Z}
\begin{split} 
&\log Z=\sum_{t=1}^L \sum_{i\neq j=1}^{N_t} (\mathfrak{g}-1)\text{Li}_1 \! \left(e^{i(u_j^{(t)}-u_i^{(t)})} \right)+\sum_{t=1}^L \sum_{i , j=1}^{N_t} \left(Z_A(u_j^{(t)} -u_i^{(t)})+Z_H(u_j^{(t)} -u_i^{(t)})\right) \\
& -\frac{1}{2}\sum_{t=1}^{L-1}\left[\sum_{i,j=1}^{N_{t+1}}Z_H(u_j^{(t+1)}-u_i^{(t+1)})-2\sum_{j=1}^{N_{t+1}}\sum_{i=1}^{N_{t}}Z_H(u_j^{(t+1)}-u_i^{(t)})+\sum_{i,j=1}^{N_{t}}Z_H(u_j^{(t)}-u_i^{(t)}) \right] \\
&-\frac{1}{2}\sum_{t \in \{ 1, L \}} \sum_{j=1}^{N_{t}}\sum_{i=1}^{N_{t}}Z_H(u_j^{(t)}-u_i^{(t)}) + k_L \sum_{i=1}^{N_L} Z_H (u_i^{(L)}) \ .
\end{split}
\end{equation}
Substituting $u=iLx$, we now consider the large $L$ limit of this expression, using the results in Appendix \ref{Appendix_index}. The procedure retraces the one used for the twisted superpotential and we will not explicitly repeat it here but we report all the details in Appendix \ref{Appendix_index}. We only note that the first line of \eqref{trick_for_Z} produces a non local term
\begin{equation}
\label{linear_z}
i\sum_{j>i}\left(u_i^{(t)} -u_j^{(t)}+\frac{1}{2}\pi \right) \left(2\mathfrak{n}_h+\mathfrak{n}_v-2(1-\mathfrak{g})\right) \ ,
\end{equation}
which vanishes imposing the condition \eqref{cond_fluxes}. This is completely analogous to what found for the free energy and the twisted superpotential when we used the conditions on R-charges and chemical potentials to make the non local terms vanish. From the other lines we get long-range contributions and, all in all, we can recast $\log \abs{Z}$ as
\begin{equation}
\label{logZ}
\begin{split}
&\log \abs{Z}= \int_0^1 dz \int dx \Bigg[ \varrho(z,x)^2\tilde{Z}_0(\Delta,\mathfrak{n}) -\frac{1}{2}\tilde{Z}_H (\mathfrak{n})\partial_z \varrho(z,x)\int dx' \partial_z \varrho(z,x')\abs{x-x'}\Bigg]  \\
& -\frac{L}{2} \sum_{z \in \{ 0,1 \} } \int dx dx' \varrho(z,x) \varrho(z,x')\tilde{Z}_H(\mathfrak{n})\abs{x-x'}+L k_L \int dx  \varrho (1,x) \tilde{Z}_H(\mathfrak{n})\abs{x}
\end{split}
\end{equation}
with 
\begin{align} 
\tilde{Z}_0(\Delta,\mathfrak{n}) &= \Delta(\pi-\frac{3}{2}\Delta)\mathfrak{n}_v +2\Delta(\Delta-\pi)(1-\mathfrak{g}) \ , \\
\tilde{Z}_H(\mathfrak{n}) &=-(-\mathfrak{n}_h+1-\mathfrak{g})=-\frac{\mathfrak{n}_v}{2} \ .
\end{align}

\subsection{Evaluation of the twisted index}
The last step is to evaluate  \eqref{logZ} in the saddle point configuration \eqref{density2}. For future convenience, we recall that this configuration satisfies
\begin{equation}
\label{saddle_cond}
\int dx' \  \partial_z^2 \varrho_s(z,x')\abs{x-x'}=-\varrho_s(z,x)2\Delta^2  \ .
\end{equation}
Using the boundary conditions \eqref{bound_W},  the explicit terms form the boundary $z \in \{0,1\}$ and the contributions from the flavours in the last node disappear. Integrating by parts the remaining expression we find
\begin{equation}
\begin{split}
\log \abs{Z}= \int_0^1 dz & \int dx  \varrho_s(z,x) \left( \varrho_s(z,x)  \tilde{Z}_0(\Delta,\mathfrak{n})+\tilde{Z}_H(\mathfrak{n})\frac{1}{2}\int dx' \partial_z^2 \varrho_s(z,x')  \abs{x-x'} \right)  \\ 
&- \tilde{Z}_H(\mathfrak{n})\frac{1}{2}\int dx dx'\left(\varrho_s(z,x) \partial_z \varrho_s(z,x')\right|^{z=1}_{z=0}\abs{x-x'} \ . 
\end{split}
\end{equation}
Substituting $\tilde{Z}_0$ and $\tilde{Z}_H$
\begin{equation}
\begin{split}
& \log \abs{Z}= \int_0^1 dz \Bigg[ \int dx \varrho_s(z,x)^2  \left(\Delta(\pi-\frac{3}{2}\Delta)\mathfrak{n}_v +2\Delta(\Delta-\pi)(1-\mathfrak{g}) \right)  \\
& \ -\frac{\mathfrak{n}_v}{4}\int dx dx' \varrho_s(z,x) \partial_z^2 \varrho_s(z,x')\abs{x-x'} \Bigg] +\frac{\mathfrak{n}_v}{4}\int dx dx'\left(\varrho_s(z,x) \partial_z \varrho_s(z,x')\right|^{z=1}_{z=0} \abs{x-x'}
\end{split}
\end{equation}
and using the saddle point condition \eqref{saddle_cond} together with the relation $2\mathfrak{n}_h+\mathfrak{n}_v=2(1-\mathfrak{g})$, we obtain
\begin{equation}
\label{integrals_Z}
\begin{split}
 \log \abs{Z}= 2\Delta(\Delta-\pi)\mathfrak{n}_h&\int_0^1 dz  \int dx \varrho_s(z,x)^2 +\frac{\mathfrak{n}_v}{4}\int dx dx'\left(\varrho_s(z,x) \partial_z \varrho_s(z,x')\right|^{z=1}_{z=0} \abs{x-x'} \ .
\end{split}
\end{equation}
The second integral is the one found in the computation of the free energy, equation \eqref{inte_useful}. After introducing an appropriate cut-off, it gives 
\begin{equation}
\label{Z_contr_1}
\frac{\mathfrak{n}_v}{4}\int dx dx'\left(\varrho_s(z,x) \partial_z \varrho_s(z,x')\right|^{z=1}_{z=0} \abs{x-x'}=- \frac{\Delta}{2\pi}\mathfrak{n}_v k_L^2\log L+\dots \ .
\end{equation}
Consider now the first integral in \eqref{integrals_Z}. Using that the density is even in $x$ and introducing a cut-off $\beta/L$ with $\beta$ arbitrary, we write it as
\begin{equation}
\label{Z_contr_2-beg}
\begin{split}
&4\Delta(\Delta-\pi)\mathfrak{n}_h \int_0^1 dz \int_{\frac{\beta}{L}} dx \left(\frac{k_L \sin(\pi z)}{2\Delta(\cosh(\frac{\pi}{\Delta}x)+\cos(\pi z))}\right)^2 = \\ 
=& (\Delta-\pi)\mathfrak{n}_h \int_0^1 dz \int_{\frac{\beta'}{L}} dx' \left(\frac{k_L \sin(\pi z)}{(\cosh(\pi x')+\cos(\pi z))}\right)^2 =  \\
= &(\Delta-\pi)\mathfrak{n}_h k_L^2 \int_{\frac{\beta'}{L}}dx'\left(\coth(\pi x')-1 \right)
\end{split}
\end{equation}
where we used the change of variable $x'=x/\Delta$ and performed the $z$ integral (using e.g.\! Mathematica). Considering the leading order in $L$, we find
\begin{equation}
\label{Z_contr_2}
(\Delta-\pi)\mathfrak{n}_h k_L^2 \int_{\frac{\beta'}{L}}dx' \left(\coth(\pi x')-1 \right)= \frac{(\Delta-\pi)}{\pi}\mathfrak{n}_h k_L^2 \log L + \dots \ .
\end{equation}
As promised, the determinant contribution in \eqref{det_contr} is actually subleading if compared with \eqref{Z_contr_1} and \eqref{Z_contr_2}, which represent the leading contribution to the index. Putting all together, we finally find at the leading order ($L=N$ and $k_L=N$)
\begin{equation}
\begin{split}
\label{value_index}
\log Z_{\Sigma_{\mathfrak{g}}\times S^1}(\Delta,\mathfrak{n})=N^2\log N \Bigg[\frac{(\Delta-\pi)}{\pi}\mathfrak{n}_h- \frac{\Delta}{2\pi}\mathfrak{n}_v\Bigg]= N^2\log N \left[\frac{(2\Delta-\pi)}{\pi}\mathfrak{n}_h- \frac{\Delta}{\pi}(1-\mathfrak{g}) \right]\ .
\end{split}
\end{equation}
Comparing \eqref{value_bethe} and \eqref{value_index}, it is easy to check that the index theorem \eqref{index-theorem} is actually verified\footnote{Note that the proof of the index theorem in \cite{Hosseini:2016tor} doesn't directly apply to the $T[SU(N)]$ case since eq. (5.3) of \cite{Hosseini:2016tor} is not satisfied by the on-shell twisted superpotential \eqref{value_bethe}.}. As a consistency check of this result, we can evaluate \eqref{value_index} on the universal twist
\begin{equation}
\label{universal_twist2}
\bar{\mathfrak{n}}_h=\frac{\bar{\Delta}}{\pi}(1-\mathfrak{g}) \ , \qquad \bar{\Delta}=\frac{\pi}{2} \ .
\end{equation}
Substituting in \eqref{value_index}, we obtain
\begin{equation}
\label{finale_1}
\log Z_{\Sigma_{\mathfrak{g}}\times S^1}(\bar{\Delta},\bar{\mathfrak{n}}) =(\mathfrak{g}-1)\frac{1}{2}N^2 \log N
\end{equation}
and
\begin{equation}
\label{finale_2}
\log Z_{\Sigma_{\mathfrak{g}}\times S^1}(\bar{\Delta},\bar{\mathfrak{n}})= (\mathfrak{g}-1) \mathcal{F}_{S^3}\left(\frac{\bar{\Delta}}{\pi} \right)
\end{equation}
as expected from the supegravity analysis of \cite{Azzurli:2017kxo}.

\section{Conclusions}
In this work, we have computed the topologically twisted index for the $T[SU(N)]$ theory, in the large $N$ limit. In particular, we focused on the case of the universal topological twist, obtained with the choice of magnetic fluxes and chemical potentials given in \eqref{universal_twist2}. The expression we found correctly reproduces the entropy of the universal black hole \citep{Azzurli:2017kxo}, satisfying \eqref{finale_1} and \eqref{finale_2}.

As an intermediate step, we also provided the computation of the free energy on the three-sphere at large $N$, turning on an arbitrary R-charge. A natural idea is then to apply the same procedure to compute the free energy of other three dimensional theories described by long linear quivers \cite{Coccia:2020wtk}. However, our discussion relies on a saddle point approximation. As we briefly mentioned in Section \ref{Section_free}, we were able to use this approximation  because the leading order of the free energy of $T[SU(N)]$, when masses and FI parameters are turned off, scales as $N^2 \log N$ and not as $N^2$. Indeed, the localization procedure gives an integral in $N^2$ variables and, in principle, all the terms in the expansion around the saddle point could contribute to this order in the free energy. This should be kept in mind before trying to apply the procedure to other theories described by long linear quivers.
Moreover, it is interesting to observe that the saddle point configuration for the three dimensional $T[SU(N)]$ theory is the same found in \cite{Uhlemann:2019ypp} for the five dimensional $T_N$ theory, up to a rescaling. Hence, the formalism applied in this paper could be used to investigate relations between three and five dimensional quiver theories \cite{Coccia:2020wtk}.

With our discussion, we applied the method of \cite{Uhlemann:2019ypp}, proposed for the computation of free energy on five-spheres, to the three dimensional case. As future directions of research, it would be then natural to study what happens in other dimensions. Moreover, as recently done in \cite{Uhlemann:2020bek} for Wilson loops, one could also try to exploit the expressions for the saddle point configurations to compute other quantities in the field theory side, comparing the results with holographic predictions.

\acknowledgments I am deeply grateful to Alberto Zaffaroni for suggesting me this project, for his continued help and for many comments on the draft. I would also like to thank Ivan Garozzo, Andrea Grigoletto, Gabriele Lo Monaco and Matteo Sacchi for many useful discussions and clarifications. I am supported by the INFN and by the MIUR-PRIN contract 2017CC72MK003.

\appendix
\section{Polylogarithms}
For ease of reading, we here recall the definition of polylogarithms  
\begin{equation}
\label{def_poly}
\text{Li}_s (z)= \sum_{k=1}^\infty \frac{z^k}{k^s}
\end{equation}
together with some useful properties
\begin{equation}
\label{polylog_prop}
\begin{split}
\Li_0(e^{iu})+\Li_0(e^{-iu})&=-1  \ ,\\
\Li_1(e^{iu})-\Li_1(e^{-iu})&=-iu+i\pi \ ,\\
\Li_2(e^{iu})+\Li_2(e^{-iu})&=\frac{u^2}{2}-\pi u+\frac{\pi^2}{3} \ ,\\
\Li_3(e^{iu})-\Li_3(e^{-iu})&=\frac{i}{6}u^3-i\frac{\pi}{2}u^2+i\frac{\pi^2}{3}u \ ,
\end{split}
\end{equation}
where we assumed $0<\mathds{R}\text{e}(u)<2 \pi$. Relations in the region $-2\pi<\mathds{R}\text{e}(u)<0$ can be found sending $u \to -u$. We also define the functions:
\begin{equation}
\label{gs}
g_{+}(u)=\frac{u^3}{6}- \frac{\pi}{2}u^2+\frac{\pi^2}{3}u \ , \qquad \qquad g'_+ (u)=\frac{u^2}{2}- \pi u +\frac{\pi^2}{3} \ .
\end{equation}
\section{Formulas for the computation of the free energy on $S^3$}
\label{Appenix_free}
As argued in Section \ref{Section_free}, the free energy on the three-sphere for $T[SU(N)]$ can be written in the form
\begin{equation}
\label{rep_free}
\begin{split}
& \mathcal{F}=  L \int_0^1 dz \int d\lambda d\lambda' N(z)^2 \rho(z,\lambda) \rho(z,\lambda')\left(F_V\left(\lambda-\lambda'\right)+F_H\left(\lambda-\lambda'\right)\right) \\
& -L \int_0^1 dz \int d\lambda d\lambda' \frac{1}{2L^{2}}\partial_z \left(N(z) \rho(z,\lambda)\right)\partial_z \left(N(z) \rho(z,\lambda') \right) F_H\left(\lambda-\lambda'\right)  \\
& -\frac{1}{2} \sum_{z \in \{ 0,1 \} } \int d\lambda d\lambda' \ N(z)^2\rho(z,\lambda) \rho(z,\lambda')F_H\left(\lambda-\lambda' \right)+k_L N(1) \int d\lambda  \rho (1,\lambda) F_H\left(\lambda\right) \ .
\end{split}
\end{equation}
However, we now show how it is possible to simplify this expression assuming the scaling 
\begin{equation}
\lambda=L^{\alpha} x
\end{equation}
with $\alpha>0 $ and in the large $L$ limit. 
\subsection*{Bi-fundamentals contribution} We start considering integrals in \eqref{rep_free} containing $F_H$ (see \cite{Jain:2019lqb,Jafferis:2011zi,Martelli:2011qj,Cheon:2011vi,Amariti:2019pky} for related computations). Consider for example
\begin{equation}
\label{hyper-contr}
\begin{split}
\int dx dy &\ \varrho(z,x)\varrho(z,y)  \ F_H (L^\alpha(x-y))= \\
& =-\int dx dy \ \varrho(z,x)\varrho(z,y) \ \left[\ell\left(1-r+i L^\alpha(x-y)\right)+\ell\left(1-r-iL^\alpha(x-y)\right)\right] \ ,
\end{split}
\end{equation}
where we used the rescaled density $dx\varrho(z,x)=d\lambda N(z)\rho(z,\lambda)$ and the definition \eqref{Fv-Fh} of $F_H$. We also recall that
\begin{equation}
\ell(z)=-z\log \left(1-e^{2\pi i z}\right)+\frac{i}{2}\left(\pi z^2+\frac{1}{\pi}\text{Li}_2\left(e^{2\pi i z}\right) \right)-\frac{i\pi}{12} \ .
\end{equation}
It is convenient to separately consider the different contributions inside $\ell(z)$, starting from the terms involving $-z\log(1-\exp[2\pi i z])$, i.e.
\begin{equation}
\label{zlog_contr}
\begin{split}
-\sum_{k=1}^\infty \frac{1}{k}\int dx dy  \ & \varrho(z,x)\varrho(z,y)\Bigg[(1-r+iL^\alpha(x-y))\left(e^{2\pi k( -i r- L^\alpha (x-y))}\right) \\
&+ (1-r-iL^\alpha(x-y))\left(e^{2\pi k(- i r+ L^\alpha (x-y))}\right) \Bigg] \ ,
\end{split}
\end{equation}
where we noted that
\begin{equation}
-\log(1-z)=\Li_1(z)=\sum_{k=1}^\infty \frac{z^k}{k} \ .
\end{equation}
Integrating by parts, we can obtain the first terms of the large $L$ expansion for the previous integral. Explicitly, consider, in the region $x>y$, the integral 
\begin{equation}
\begin{split}
& -\int_{-\infty}^x dy  \ \varrho(z,y)\Bigg[(1-r+iL^\alpha(x-y))\left(e^{2\pi k( -i r- L^\alpha (x-y))}\right) \ ,
\end{split}
\end{equation}
which, after an integration by parts, becomes
\begin{equation}
\begin{split}
\frac{\varrho(z,x)}{L^\alpha} & \left(\frac{(r-1)}{2 \pi k}-\frac{i}{4\pi^2 k^2 }\right)e^{-2 \pi i k r} \\
& + \frac{1}{L^\alpha }\int_{-\infty}^x \partial_y \varrho(z,y) \  e^{2 \pi k(-i r-L^\alpha (x-y)}\frac{\left(i+2 k \pi (1-r+iL^\alpha(x-y)) \right)}{4 k^2 \pi^2 } \ .
\end{split}
\end{equation}
If one keep integrating by parts the second line, obtains an expansion in $1/L^\alpha$. Saving only the leading order, then, we write the first line of \eqref{zlog_contr} in the region $x>y$ as
\begin{equation}
\label{first1}
\begin{split}
&-\sum_{k=1}^\infty \frac{1}{k}\int dx \ \varrho(z,x) \int_{-\infty}^x dy  \ \varrho(z,y)(1-r+iL^\alpha(x-y))\left(e^{2\pi k( -i r- L^\alpha (x-y))}\right)= \\
=&\sum_{k=1}^\infty \frac{1}{k}\Bigg[ \int dx \  \varrho(z,x)^2 \left(\frac{(r-1)}{2 \pi k L^\alpha}e^{-2 \pi i k r} -\frac{i}{4 \pi^2 k^2 L^\alpha}e^{-2 \pi i k r}\right) \Bigg]= \\
=&\int dx \ \varrho(z,x)^2 \left(\frac{(r-1)}{2 \pi L^\alpha}\Li_2(e^{-2 \pi i r})-\frac{i}{4  \pi^2 L^\alpha}\Li_3(e^{-2 \pi i r})\right) \ .
\end{split}
\end{equation}
If we try to apply the same procedure in the region $x<y$, we encounter divergences in the integration by parts. Fortunately, we can use \eqref{polylog_prop} to invert the sign in the exponential and avoid divergences. With this procedure, however, we also obtain a non local term:
\begin{equation}
\label{first2}
\begin{split}
\int dx \int_x^\infty dy \  & \varrho(z,x)\varrho(z,y)   (1-r+iL^\alpha(x-y))\left(-2\pi i  r-2\pi L^\alpha(x-y)+i\pi\right) \\
& +\int dx \ \varrho(z,x)^2 \left(\frac{(r-1)}{2 \pi L^\alpha}\Li_2(e^{2 \pi i r})+\frac{i}{4 \pi^2 L^\alpha}\Li_3(e^{2 \pi i r})\right) \ .
\end{split}
\end{equation}
Summing \eqref{first1} and \eqref{first2}, we find for the first line of \eqref{zlog_contr}
\begin{equation}
\begin{split}
\int dx \int_x^\infty dy \ \varrho(z,x)\varrho(z,y) \ & (1-r+iL^\alpha(x-y))\left(-2\pi i r-2 \pi L^\alpha(x-y)+i\pi\right) \\
& +\int dx \ \varrho(z,x)^2 \left[ \frac{(r-1)}{2 \pi L^\alpha}g'_+(2 \pi r)-\frac{g_+(2 \pi r)}{4 \pi^2 L^\alpha}\right]
\end{split}
\end{equation}
with $g_+(u)$ defined in \eqref{gs}. The second line in \eqref{zlog_contr} can be computed in the same way and gives
\begin{equation}
\begin{split}
\int dx \int_{-\infty}^x dy \ \varrho(z,x)\varrho(z,y) \ & (1-r-iL^\alpha(x-y))\left(-2\pi i r+2 \pi L^\alpha(x-y)+i\pi\right) \\
&  +\int dx \ \varrho(z,x)^2 \left[ \frac{(r-1)}{2 \pi L^\alpha}g'_+(2 \pi r)-\frac{g_+(2 \pi r)}{4 \pi^2 L^\alpha}\right] \ .
\end{split}
\end{equation}
All together, the leading contribution in $L$ from \eqref{zlog_contr} is
\begin{equation}
\label{zlog1}
\begin{split}
\int dx \int dy \ \varrho(z,x)\varrho(z,y) \ & (1-r-iL^\alpha \abs{x-y})\left(-2\pi i r+2 \pi L^\alpha\abs{x-y}+i\pi\right) \\
&  +\int dx \ \varrho(z,x)^2 \left[ \frac{(r-1)}{\pi L^\alpha}g'_+(2 \pi r)-\frac{g_+(2 \pi r)}{2 \pi^2 L^\alpha}\right] \ .
\end{split}
\end{equation}

The computation of dilogarithms contributions in \eqref{hyper-contr}  is very similar in the procedure. The result is
\begin{equation}
\label{zlog2}
\begin{split}
-\frac{i}{2\pi}\int dx dy \ \varrho(z,x) \varrho(z,y) \ & \left[ \frac{\left(-2\pi i L^\alpha \abs{x-y}-2\pi r \right)^2}{2}+\pi\left(-2\pi i L^\alpha \abs{x-y}-2\pi r\right)+\frac{\pi^2}{3}\right]  \\
&-\frac{1}{2\pi^2 L^\alpha}\int dx \varrho(z,x)^2 g_+(2\pi r) \ .
\end{split}
\end{equation}
Expressions \eqref{zlog1} and \eqref{zlog2}, together with the remaining quadratic and constant terms in $\ell(z)$, finally give the leading order contribution 
\begin{equation}
\label{FH_contribution}
\begin{split}
\int & dx dy \ \varrho(z,x)\varrho(z,y)  \ F_H\!(L^\alpha(x-y))= \\
&=\frac{\pi}{3L^\alpha}(r-1)(1+2r(r-2))\int dx \ \varrho(z,x)^2+2\pi L^\alpha(1-r) \int dx dy \ \varrho(z,x) \varrho(z,y) \abs{x-y} \ .
\end{split}
\end{equation}
A completely analogous argument can be used for the other integrals involving $F_H$.

\subsection*{Adjoint and vector contribution}
Next, we need to consider the term in \eqref{rep_free} involving $F_V$, which represents the adjoint and the vector contributions. The former can be thought as bi-fundamental connecting the same gauge group and we can then use the result in \eqref{FH_contribution}, with an overall factor $1/2$ (see \eqref{Fv-Fh})
\begin{equation}
\begin{split}
&-\frac{1}{2}\int dx dy \ \varrho(z,x)\varrho(z,y) \ \left[\ell\left(1-\tilde{r}+i L^{\alpha}(x-y)\right)+\ell\left(1-\tilde{r}-iL^{\alpha}(x-y)\right)\right] = \\
&=\frac{\pi}{6L^{\alpha}}(\tilde{r}-1)(1+2\tilde{r}(\tilde{r}-2))\int dx \ \varrho(z,x)^2+\pi L^{\alpha}(1-\tilde{r}) \int dx dy \ \varrho(z,x) \varrho(z,y) \abs{x-y} \ .
\end{split}
\end{equation}
The vector contribution is, instead, given by the integral of (see again \eqref{Fv-Fh})
\begin{equation}
-\frac{1}{2}\log\left(4 \sinh^2(L^\alpha \pi(x-y)) \right)=-\pi L^\alpha \abs{x-y}-\log\left(1-e^{-2 \pi L^\alpha \abs{x-y}} \right) \ ,
\end{equation}
which in the planar limit becomes
\begin{equation}
\begin{split}
-\frac{1}{2}\int dx dy & \ \varrho(z,x) \varrho(z,y)  \log (4\sinh^2(L^\alpha \pi (x-y)))= \\ &=  -L^\alpha \pi\int dx dy \ \varrho(z,x) \varrho(z,y)\abs{x-y} +\frac{\pi}{6 L^\alpha}\int dx \ \varrho(z,x)^2 \ .
\end{split}
\end{equation}
Here, we integrated by parts the term with the logarithm as we did before, throwing away subleading orders in $L^\alpha$.
So, all together
\begin{equation}
\label{FV_contribution}
\begin{split}
\int &dx dy \ \varrho(z,x)\varrho(z,y)  \ F_V (L^\alpha (x-y))= \\
&=-\pi \tilde{r} L^{\alpha} \int dx dy \ \varrho(z,x) \varrho(z,y)\abs{x-y} +\frac{\pi}{6L^{\alpha}}\left[(\tilde{r}-1)(1+2\tilde{r}(\tilde{r}-2))+1 \right]\int dx \ \varrho(z,x)^2 \ .
\end{split}
\end{equation}

We conclude this appendix with a couple of important observations. In the expression \eqref{rep_free}, the combination $F_0=F_V+F_H$ appears. Summing together \eqref{FH_contribution} and \eqref{FV_contribution} we see that the non local term is zero when $2r+\tilde{r}=2$, condition required by the superpotential. The term in the integral involving $F_0$ is then local in the density and we can write
\begin{equation}
\int dx dy \ \varrho(z,x)\varrho(z,y) F_0(L^{\alpha}(x-y))= \frac{2\pi (1-r)r^2}{L^\alpha}\int dx \varrho(z,x)^2 \ .
\end{equation}
Conversely, all the other terms in \eqref{rep_free} in which only $F_H$ appears are non local and we will only keep the leading long-range force contribution from \eqref{FH_contribution}.

\section{Formulas for the computation of the twisted index}
\label{Appendix_index}
In the computation of the index, we need to evaluate
\begin{equation}
\log Z_{pert} \Big\lvert_{\mathfrak{m}=0} =  \log\left(\prod_{\alpha \in G}(1-\lambda^{\alpha})^{1-\mathfrak{g}} \prod_I\prod_{\rho_I \in \mathfrak{R}_I} \left( \frac{\lambda^{\rho_I/2}y_I^{1/2}}{1-\lambda^{\rho_I}y_I}\right)^{-\mathfrak{n}_I+1-\mathfrak{g}} \right)
\end{equation}
where the product over the roots $\alpha$ of $G$ is the contribution from the $\mathcal{N}=2$ vector multiplet and the other products denote the contribution of the chiral multiplets in the theory. Computations of this appendix can be compared with those of \cite{Hosseini:2016tor,Jain:2019lqb}. We also recall that, in the upcoming computations, overall phases in the index can be neglected, since we will be interested in $\log \abs{Z}$.

\subsection*{Bi-fundamentals contribution} Let us start considering the logarithm of bi-fundamentals contribution, made of two fields connecting two  adjacent nodes labelled by $t$ and $t+1$. Associated with them we have a magnetic flux $\mathfrak{n}_h$ and a fugacity $y_h=e^{i\Delta}$, equal for both the fields. Hence, we have
\begin{equation}
\begin{split}
&\sum_{i=1}^{N_t}\sum_{j=1}^{N_{t+1}} Z_H(u_j^{(t+1)}-u_i^{(t)}) \equiv \log \prod_{i=1}^{N_t} \prod_{j=1}^{N_{t+1}} \left[-\frac{\lambda_j^{(t+1)}}{\lambda_i^{(t)}} \frac{1}{\left(1-\frac{\lambda_j^{(t+1)}}{\lambda_i^{(t)}}y_h^{(-1)}\right)\left(1-\frac{\lambda_j^{(t+1)}}{\lambda_i^{(t)}}y_h\right)}\right]^{-\mathfrak{n}_h+1-\mathfrak{g}} \ ,
\end{split}
\end{equation}
which we rewrite
\begin{equation}
\label{bifund-contr-Z}
\begin{split}
&\sum_{i=1}^{N_t}\sum_{j=1}^{N_{t+1}} Z_H(u_j^{(t+1)}-u_i^{(t)}) = \\ & = (-\mathfrak{n}_h+1-\mathfrak{g}) \sum_{i=1}^{N_t} \sum_{j=1}^{N_{t+1}} \Bigg[\text{Li}_1\left(e^{i\left(u_j^{(t+1)}-u_i^{(t)}+\Delta \right)}\right)+ \text{Li}_1 \left( e^{i\left(u_j^{(t+1)}-u_i^{(t)}-\Delta \right)} \right)  \\
& \qquad \qquad \qquad \qquad \qquad \qquad \qquad \qquad \qquad \qquad  + i\left(u_j^{(t+1)}-u_i^{(t)} \right)-i\pi \Bigg] 
\end{split}
\end{equation}
using $\lambda=e^{iu}$ and $\Li_1(z)=-\log(1-z)$.  Now, consider the first logarithm in  \eqref{bifund-contr-Z}.  With the assumption $u=iLx$, in the large $L$ limit the region $j>i$ gives
\begin{equation}
\label{specimen}
\begin{split}
\int dx \ \varrho_t(x) \int_x^{\infty} dx' \ \varrho_{t+1}(x') \ \text{Li}_1 \left( e^{-L(x'-x)+i\Delta} \right) \ .
\end{split}
\end{equation}
This integral is analogous to those found in Appendix \ref{Appenix_free}. As we did in Appendix \ref{Appenix_free} for the free energy, we apply the definition \eqref{def_poly} and integrate by parts to obtain the expansion in $1/L$. We find
\begin{equation}
\int dx \ \varrho_t(x)  \int_x dx' \varrho_{t+1}(x') \ \text{Li}_1 \left( e^{-L(x'-x)+i\Delta} \right)= \frac{1}{L}\int dx \ \varrho_{t}(x)\varrho_{t+1}(x) \text{Li}_2\left(e^{i\Delta} \right)+ \dots
\end{equation}
When $j<i$ instead,  we need to invert the integrand to avoid divergences in the procedure of integration by parts. This can be done using the properties \eqref{polylog_prop}. All together, from the first logarithm in \eqref{bifund-contr-Z}, at the leading order in $L$ we have
\begin{equation}
\label{local-bif-Z}
\begin{split}
\frac{1}{L}\int dx \ \varrho_{t}(x)\varrho_{t+1}(x)\left( \text{Li}_2\left(e^{i\Delta} \right)+\text{Li}_2\left(e^{-i\Delta} \right) \right)=\frac{1}{L}\int dx \ \varrho_{t}(x)\varrho_{t+1}(x)g_+'(\Delta)
\end{split}
\end{equation}
plus non local terms from the inversion formula
\begin{equation}
\begin{split}
 -i \sum_{i>j}\left( u_j^{(t+1)}-u_i^{(t)} +\Delta-\pi \right) \ .
\end{split}
\end{equation}
The second logarithm in \eqref{bifund-contr-Z} can be treated analogously and gives the same local term \eqref{local-bif-Z}, plus 
\begin{equation}
\begin{split}
 -i \sum_{i>j}\left( u_j^{(t+1)}-u_i^{(t)} -\Delta+\pi \right) \ .
\end{split}
\end{equation}
Notice that we obtained this contribution applying relations \eqref{polylog_prop} in the region $-2\pi<\mathds{R}\text{e}(u)<0$. In the end, considering all the terms in \eqref{bifund-contr-Z}, we have
\begin{equation}
\label{ZH}
\begin{split}
\frac{2}{L}(-\mathfrak{n}_h+1-\mathfrak{g})\int dx \ \varrho_t(x)\varrho_{t+1}(x)  g_+'(\Delta) 
\end{split}
\end{equation}
and the long-range term
\begin{equation}
\label{ZH-nonloc}
-i(-\mathfrak{n}_h+1-\mathfrak{g})\sum_{i \neq j}\left[ \left(u_j^{(t+1)} -u_i^{(t)} \right) \text{sign}(i-j)+\pi \right] \ .
\end{equation} 
For the moment, it is convenient not to keep the continuous limit of this sum.

\subsection*{Adjoint contribution} Adjoint terms can be identified with bi-fundamentals connecting the same node. Calling $\mathfrak{n}_v$ the magnetic flux  and $\tilde{\Delta}$ the chemical potential 
\begin{equation}
\begin{split}
\sum_{i,j=1}^{N_t} Z_V(u_j^{(t)}-u_i^{(t)})\equiv \frac{1}{2}(-\mathfrak{n}_v+1-\mathfrak{g}) \sum_{i,j=1}^{N_t} \Bigg[\text{Li}_1\left(e^{i\left(u_j^{(t)}-u_i^{(t)}+\tilde{\Delta} \right)}\right)&+\text{Li}_1 \left( e^{i\left(u_j^{(t)}-u_i^{(t)}-\tilde{\Delta} \right)} \right)  \\
& +i\left(u_j^{(t)}-u_i^{(t)} \right)-i\pi \Bigg] \ .
\end{split}
\end{equation}
Hence, we can use the results \eqref{ZH}, \eqref{ZH-nonloc} and conclude that, in the continuous limit, the contribution from each adjoint field is
\begin{equation}
\label{ZV1}
\begin{split}
\frac{1}{L}(-\mathfrak{n}_v+1-\mathfrak{g}) \int dx \ \varrho_t(x)^2 g_+'(\tilde{\Delta}) 
\end{split}
\end{equation}
plus the non local term
\begin{equation}
-\frac{i}{2}(-\mathfrak{n}_v+1-\mathfrak{g})\sum_{i \neq j}^{N_t}\left[ \left(u_j^{(t)} -u_i^{(t)} \right) \text{sign}(i-j)+\pi \right] \ .
\end{equation}

\subsection*{(Anti-)fundamental contribution} For the (anti-)fundamental contribution we have
\begin{equation}
\begin{split}
&(-\mathfrak{n}_h+1-\mathfrak{g}) \sum_{i=1}^{N_t}\log \left(-\frac{x_i^{(t)}}{\left(1-x_i^{(t)}y_h\right)\left(1-x_i^{(t)}y_h^{(-1)}\right)} \right)= \\
=& (-\mathfrak{n}_h+1-\mathfrak{g})\sum_{i=1}^{N_t} \left[\text{Li}_1\left( e^{i(u_i^{(t)}+\Delta)} \right)+\text{Li}_1\left( e^{i(u_k^{(t)}-\Delta)}\right)+i u_i^{(t)}-i\pi\right] \ ,
\end{split}
\end{equation}
which, in the continuous limit, has the leading order
\begin{equation}
\label{index_fund}
-(-\mathfrak{n}_h+1-\mathfrak{g})L\int dx \ \varrho_t(x) \abs{x} \ .
\end{equation}
Adding to our discussion other fluxes associated to fundamentals and anti-fundamentals, call them $\mathfrak{n}_f$ and $\mathfrak{n}_f'$ respectively, we would obtain a subleading contribution due to the condition $\mathfrak{n}_f=-\mathfrak{n}_f'$ imposed by the superpotential.

\subsection*{Gauge vector contribution}
To conclude, we consider the term coming from the vector multiplet in the $t^{\text{th}}$ node 
\begin{equation}
 \log \prod_{i \neq j}^{N_t} \left(1-\frac{x_i^{(t)}}{x_j^{(t)}} \right)^{1-\mathfrak{g}}  \ ,
\end{equation}
which we rewrite
\begin{equation}
\begin{split}
&(1-\mathfrak{g})\left[ \sum_{i > j}\log\! \left(1-e^{i(u_i^{(t)}-u_j^{(t)})} \right)+ \sum_{j > i}\log\! \left(1-e^{i(u_i^{(t)}-u_j^{(t)})} \right) \right]= \\
&= (1-\mathfrak{g})\left[ \sum_{i > j}\log\! \left(1-e^{i(u_i^{(t)}-u_j^{(t)})} \right)+ \sum_{j > i}\left(\log\! \left(1-e^{i(u_j^{(t)}-u_i^{(t)})} \right)+i(u_i^{(t)}-u_j^{(t)})+i\pi \right) \right]= \\
& = (\mathfrak{g}-1)\left(2\sum_{i > j}\text{Li}_1 \left(e^{i(u_i^{(t)}-u_j^{(t)})} \right)- i\sum_{j>i} \left(u_i^{(t)}-u_j^{(t)}+\pi\right) \right) 
\end{split}
\end{equation}
and in the large $L$ limit gives the local term
\begin{equation}
\label{ZV2}
(\mathfrak{g}-1)\frac{\pi^2}{3L}\int dx \ \varrho_t(x)^2  
\end{equation}
plus
\begin{equation}
i(1-\mathfrak{g})\sum_{j>i} \left(u_i^{(t)}-u_j^{(t)}+\pi\right) \ .
\end{equation}

\subsection*{Other manipulations}
Let us now use the previous results to compute the continuous limit of Eq.\! \eqref{trick_for_Z}. The first term we need to consider is the combination 
\begin{equation}
\label{comb-bif-adj-vec}
\sum_{t=1}^L \sum_{i\neq j=1}^{N_t} (\mathfrak{g}-1)\text{Li}_1 \! \left(e^{i(u_j^{(t)}-u_i^{(t)})} \right)+\sum_{t=1}^L \sum_{i , j=1}^{N_t} \left(Z_A(u_j^{(t)} -u_i^{(t)})+Z_H(u_j^{(t)} -u_i^{(t)})\right) \ .
\end{equation}
Using the results just found, we see that the long range contribution from this combination 
\begin{equation}
i\sum_{j>i}\left(u_i^{(t)} -u_j^{(t)}+\frac{1}{2}\pi \right) \left(2\mathfrak{n}_h+\mathfrak{n}_v-2(1-\mathfrak{g})\right)
\end{equation}
vanishes when $2 \mathfrak{n}_h+\mathfrak{n}_v=2(1-\mathfrak{g})$. Hence, \eqref{comb-bif-adj-vec} only produces a local term in the continuous limit, namely
\begin{equation}
\left(\mathfrak{n}_v\Delta(\pi-\frac{3}{2}\Delta)+(1-\mathfrak{g})2\Delta(\Delta-\pi) \right)\int dz \int dx \ \varrho(z,x)^2 \ .
\end{equation}

The second line of \eqref{trick_for_Z}, instead, is
\begin{equation}
-\frac{1}{2}\sum_{t=1}^{L-1}\left[\sum_{i,j=1}^{N_{t+1}}Z_H(u_j^{(t+1)}-u_i^{(t+1)})-2\sum_{j=1}^{N_{t+1}}\sum_{i=1}^{N_{t}}Z_H(u_j^{(t+1)}-u_i^{(t)})+\sum_{i,j=1}^{N_{t}}Z_H(u_j^{(t)}-u_i^{(t)}) \right]
\end{equation}
and in this case the long-range term doesn't disappear. In fact, this is the leading contribution which, in the large $L$ limit, becomes
\begin{equation}
\label{second_index}
\frac{1}{2}(-\mathfrak{n}_h+1-\mathfrak{g})L\int  dxdx' \abs{x-x'}\left(\varrho_{t+1}(x)-\varrho_{t}(x)\right)\left(\varrho_{t+1}(x')-\varrho_{t}(x')  \right)
\end{equation}
and introducing the variable $z=t/L$
\begin{equation}
\label{second_index}
\frac{1}{2}(-\mathfrak{n}_h+1-\mathfrak{g})L^{(-1)}\int dx dx' \abs{x-x'}\partial_z \varrho(z,x)\partial_z \varrho(z,x') \ .
\end{equation}

Finally, the last line in equation \eqref{trick_for_Z} is given by the contributions from the first and the last node
\begin{equation}
-\frac{1}{2}\sum_{t \in \{ 1, L \}} \sum_{j=1}^{N_{t}}\sum_{i=1}^{N_{t}}Z_H(u_j^{(t)}-u_i^{(t)}) + k_L \sum_{i=1}^{N_L} Z_H  (u_i^{(L)})
\end{equation}
and the leading order in the continuous limit is given by the non local term
\begin{equation}
\frac{L}{2}(-\mathfrak{n}_h+1-\mathfrak{g})\sum_{z \in \{ 0,1 \}}\int dx dx' \  \varrho(z,x)\varrho(z,x') \abs{x-x'}-k_L(-\mathfrak{n}_h+1-\mathfrak{g})L\int dx  \ \varrho(1,x) \abs{x} \ .
\end{equation}

\nocite{Aharony:2008ug}
\nocite{Herzog:2010hf}
\nocite{Pufu:2016zxm}
\nocite{Jafferis:2012iv}
\nocite{Hosseini:2017fjo}
\nocite{Hosseini:2018uzp}
\nocite{Hosseini:2018usu}
\nocite{Fluder:2019szh}
\nocite{Fluder:2018chf}
\nocite{DHoker:2007zhm}
\nocite{DHoker:2007hhe}
\nocite{Assel:2012cj}

\bibliographystyle{ytphys}

\bibliography{ref} 

\end{document}